\documentclass[12pt]{iopart}
\usepackage{latexsym}
\usepackage{amssymb}
\usepackage{epsf,epsfig}
\usepackage{amssymb}
\usepackage{iopams}
\usepackage{amstext}
\begin{document}
%%%%%%%%%%%%%%%%%%%%%%%%%%%%%%%%%%%%%%%%%%%%%%%%%%%%%%%%%%%%
\title{From aggressive driving to molecular motor traffic} 

\author{Ambarish Kunwar\dag, Andreas Schadschneider\ddag, and
Debashish Chowdhury\dag}
\address{\dag Department of Physics, Indian Institute of Technology,
Kanpur 208016, India}
\address{\ddag Institut f\"ur Theoretische Physik, Universit\"at zu
K\"oln, Z\"ulpicher Str. 77, D-50937 K\"oln, Germany} 
\date{\today}
%%%%%%%%%%%%%%%%%%%%%%%%%%%%%%%%%%%%%%%%%%%%%%%%%%%%%%%%%%%
\begin{abstract}
Motivated by recent experimental results for the step sizes of dynein 
motor proteins, we develope a cellular automata model for intra-cellular 
traffic of dynein motors incorporating special features of the  
hindrance-dependent step size of the individual motors. We begin by 
investigating the properties of the aggressive driving model (ADM), a 
simple cellular automata-based model of vehicular traffic, a unique 
feature of which is that it allows a natural extension to capture the 
essential features of dynein motor traffic. We first calculate several 
collective properties of the ADM, under both periodic and open boundary 
conditions, analytically using two different mean-field approaches as 
well as by carrying out computer simulations. Then we extend the ADM 
by incorporating the possibilities of attachment and detachment of 
motors on the track which is a common feature of a large class of motor 
proteins that are collectively referred to as cytoskeletal motors. 
The interplay of the boundary and bulk dynamics of attachment and 
detachment of the motors to the track gives rise a phase where high and 
low density phases separated by a stable domain wall coexist. We also 
compare and contrast our results with the model of Parmeggiani et. al. 
(Phys. Rev. Lett. {\bf 90}, 086601 (2003)) which can be regarded as a 
minimal model for traffic of a closely related family of motor proteins 
called kinesin. Finally, we compare the transportation efficiencies of 
dynein and kinesin motors over a range of values of the model parameters. 
 
\end{abstract}
%%%%%%%%%%%%%%%%%%%%%%%%%%%%%%%%%%%%%%%%%%%%%%%%%%%%%%%%%%%

%\keywords{Cellular automata, Aggressive driving, dynein motor, 
%Fundamental diagram}

\maketitle

%%%%%%%%%%%%%%%%%%%%%%%%%%%%%%%%%%%%%%%%%%%%%%%%%%%%%%%%%%%%%%%%%%%%%%%%%%%%%%
\section{Introduction}
%%%%%%%%%%%%%%%%%%%%%%%%%%%%%%%%%%%%%%%%%%%%%%%%%%%%%%%%%%%%%%%%%%%%%%%%%%%%%%

Molecular motors are protein molecules that drive a wide range of 
intra-cellular activities including transport of molecular cargo 
\cite{schliwa,howard}. 
There are many similarities between collective molecular motor 
transport and vehicular traffic \cite{reviews,physica}. In recent 
years non-equilibrium statistical mechanics has found unusual 
application in research on traffic flow of various different types 
of objects starting from objects as small as molecular motors to 
macroscopic objects like vehicles \cite{debch1,debch2,reviews,physica}. 
Analytical as well as numerical techniques of the statistical physics 
are being used to understand rich variety of physical phenomena 
exhibited by traffic systems. Some of these phenomena, observed 
under different circumstances, include phase transitions, criticality 
and self-organized criticality, metastability and hysteresis, 
phase-segregation,etc.

A common modeling strategy is to represent the motile objects 
(e.g., a vehicle or a molecular motor) by a self-propelled particle, 
ignoring its structural details, and then treating the traffic as 
a system of interacting particles driven far from equilibrium. 
These models belong to a class of non-equilibrium systems called 
{\it driven-diffusive lattice gases} \cite{zia,schutz,priv,marro}.
In most of these traffic models the dynamics of the particles is 
formulated using the language of {\it cellular automata} (CA) 
\cite{wolfram}. 

To our knowledge, the first model for molecular motor traffic was
formulated in 1968 in the context of collective movement of ribosomes
on messenger RNA track \cite{macdonald68,macdonald69}.  In recent
years several groups have independently developed a class of minimal
generic models for traffic of molecular motors which move on tracks
that are filamentary proteins. All these models are essentially
extensions of the totally asymmetric simple exclusion process (TASEP)
\cite{derrida,zia} which is one of the simplest models of driven
diffusive lattice gas systems.  In these models
\cite{frey,frey2,lipo1,lipo2,santen,popkov} the molecular motors are
represented by particles whereas the sites for the binding of the
motors with the tracks are represented by a one-dimensional discrete
lattice. Just as in TASEP, the motors are allowed to hop forward, with
probability $q$, provided the site in front is empty. However, unlike
TASEP, the particles can also get ``attached'' to an empty lattice
site, with probability $ \omega_A$, and ``detached'' from an occupied
site, with probability $\omega_D$ from any site except the end points.
Parmeggiani et al.\ \cite{frey} demonstrated a novel phase where low
and high density regimes, separated from each other by domain walls,
coexist. They interpreted this spatial organization as traffic jam of
molecular motors.

None of the models of molecular motor traffic mentioned above 
distinguish between kinesins and dyneins which form the two 
superfamilies of motor proteins that move on the same type of 
tracks, namely, microtubules. On the other hand, detailed 
experiments over the last two years have established that, 
in contrast to kinesins, dyneins can take steps of four 
different sizes depending on the opposing force or hindrance. 
One of the aims of this paper is to introduce a minimal model 
that distinguishes between these two features of kinesin 
and dynein motors.

In this paper we begin by investigating the aggressive driving 
model (ADM), a stochastic CA model for traffic flow 
\footnote{Originally the model was introduced in \cite{astgf97}.} 
that is closely related to the Nagel-Schreckenberg (NaSch) model 
\cite{nagel,ito1}. One of the reasons for studying this model is 
that it allows natural extensions so as to capture the essential 
features of dynein motor traffic including the unique features of 
dynein stepping (which we shall explain in section \ref{dyneinexperiment}). 
Besides, the ADM model is an interesting 
model of vehicular traffic in its own right and is also related to 
the Fukui-Ishibashi (FI) model \cite{fi}. However, in contrast to 
the FI model, it still shows spontaneous jam formation. We 
investigate the properties of the ADM by approximate analytical 
calculations as well as by computer simulations. Then, we use an 
extended version, which we refer to as the dynein traffic model 
(DTM), for a quantitative desciption of intra-cellular traffic of 
dynein motors.

The paper is organized as follows. In the next section we describe 
the ADM and the method of simulation. In section \ref{sec3} we
investigate the properties of the ADM with periodic boundary
conditions and we describe the analytical theories for calculating its
flow properties. We present a comparison of the ADM with NaSch
model at the end of section \ref{sec3}. In section \ref{sec4} we
investigate the density profiles and phase diagram of the ADM with
open boundary conditions. In section \ref{sec5} we describe the 
experimentally observed hindrance-dependence of the step sizes of 
dynein motors and introduce the dynein traffic model (DTM). We 
present the results for the DTM with periodic boundary conditions  
in section \ref{sec6} and those under open boundary conditions in  
section \ref{sec7}. Finally we summarize the main results and the 
conclusions in section \ref{sec8}.

%%%%%%%%%%%%%%%%%%%%%%%%%%%%%%%%%%%%%%%%%%%%%%%%%%%%%%%%%%%%%%%%%%%%%%%%%%%%%%
\section{The CA Model of Aggressive Driving}
\label{sec2}
%%%%%%%%%%%%%%%%%%%%%%%%%%%%%%%%%%%%%%%%%%%%%%%%%%%%%%%%%%%%%%%%%%%%%%%%%%%%%%

In the cellular automata model of aggressive driving a lane is
represented by a one-dimensional lattice. The boundary conditions may
be periodic or open. Each of the lattice sites represents a cell that
can be either empty or occupied by at most one vehicle at a time. The
speed $V$ of each vehicle can take one of the allowed integer values
$V=0,1,2,....V_{\rm max}$. Let $x_{n}$ and $V_{n}$ be the position and
speed, respectively, of the $n$th vehicle. 
%Then $d_{n} = x_{n+1}-x_{n}$ is the distance between the $n$th vehicle
%and the vehicle in front of it at time $t$.  The gap (number of empty
%cells) in front of the $n$th vehicle is given by $ G_{n}=d_{n}-1 $. 
Then we define the (distance) headway of the $n$th vehicle
at time $t$ by 
$d_{n} = x_{n+1}-x_{n}-1$, i.e.\ as the number of empty cells in front
of this car.
At each time step $t
\rightarrow t+1$ the state of all vehicles on this 1-D lattice is
updated in {\it parallel} according to the following rules: \\
%\begin{enumerate}
$I$: {\it Acceleration:} If $d_{n} \ge V_{\rm max}$ then $V_{n}
\rightarrow V_{\rm max}$ and if $d_{n} < V_{\rm max}$ then $V_{n}
\rightarrow d_{n}$, that is, 
$V_{n} = \min(V_{\rm max},d_{n})$ \\
$II$:{\it Randomization:} If $V_{n} > 0$, the speed of the $n$th
vehicle is decreased randomly by one with probability $p$; that is,
$V_{n}=\max(V_{n}-1,0)$ with probability $p$ \\
$III$:{\it Vehicle movement:} Each vehicle is moved forward so that $x_{n} 
\rightarrow x_{n}+V_{n}$. \\
%\end{enumerate}
Step $I$ reflects the tendency of drivers to drive the vehicle as fast
possible, without exceeding the maximum speed of the vehicle, and 
avoiding accidents between vehicles at the same time. Thus, if there 
is enough gap in front, vehicles in this model can accelerate to the 
maximum allowed velocity within one single timestep.This captures at 
least one type of aggressive driving and hence the name. The 
randomization in the step $II$ takes into account the different 
behavioral patterns of the individual drivers, especially 
non-deterministic acceleration and over-reaction while slowing down.

As usual, the {\it flux} is defined to be the number of vehicles 
crossing a detector site per unit time. In the context of vehicular 
traffic, the most important quantity of interest is the so-called 
{\it fundamental diagram} which depicts the dependence of flux on 
the density of vehicles. The number of empty sites in between a pair 
of vehicles is usually taken as a measure of the corresponding {\it  
distance-headway}. The {\it time-headway} is defined as the time
interval between the passage of two successive vehicles recorded by 
a detector placed at a fixed position on the highway. We have 
calculated all these characteristic quantities for the ADM and will 
present these results in the following sections. 

Before presenting the results for the ADM, we would like to compare 
and contrast it with a few other well known models of vehicular traffic. 
In the NaSch model, the calculation of the speed of a vehicle at the
next time step ($t+1$) during the acceleration stage requires the
knowledge of its speed at previous time step $t$ and its speed after
the deceleration stage depends on the available headway in front of it,
whereas, in the aggressive driving model the calculation of the speed of a
vehicle at next time step does not require any knowledge of its
velocity at previous time step and depends only on the available headway
in front of the vehicle. In contrast to the NaSch model it therefore
has no velocity memory. From now onwards, we shall refer to this
model as aggressive driving model (ADM).

This ADM differs from the Fukui-Ishibashi (FI) model \cite{fi}
at Step $II$ of the updating procedure. In the FI model the
randomization is applied only to those vehicles whose final velocities
become $V_{\rm max}$ after the acceleration stage and, therefore, the 
FI model is unrealistic for normal traffic.
Consequently, the FI model fails to capture overreactions at braking 
which are responsible for spontaneous jam formation 
(see e.g.\ \cite{altenberg}).

%%%%%%%%%%%%%%%%%%%%%%%%%%%%%%%%%%%%%%%%%%%%%%%%%%%%%%%%%%%%%%%%%%%%%%%%%%%%%%
\section{Results for ADM with periodic boundary conditions}
\label{sec3}
%%%%%%%%%%%%%%%%%%%%%%%%%%%%%%%%%%%%%%%%%%%%%%%%%%%%%%%%%%%%%%%%%%%%%%%%%%%%%%

%%%%%%%%%%%%%%%%%%%%%%%%%%%%%%%%%%%%%%%%%%%%%%%%%%%%%%%%%%%%%%%%%%%%%%%%%%%%%%
\subsection{Numerical results of computer simulations}
%%%%%%%%%%%%%%%%%%%%%%%%%%%%%%%%%%%%%%%%%%%%%%%%%%%%%%%%%%%%%%%%%%%%%%%%%%%%%%

\begin{figure}[tb]
\begin{center}
  \includegraphics[angle=-90,width=0.48\columnwidth]{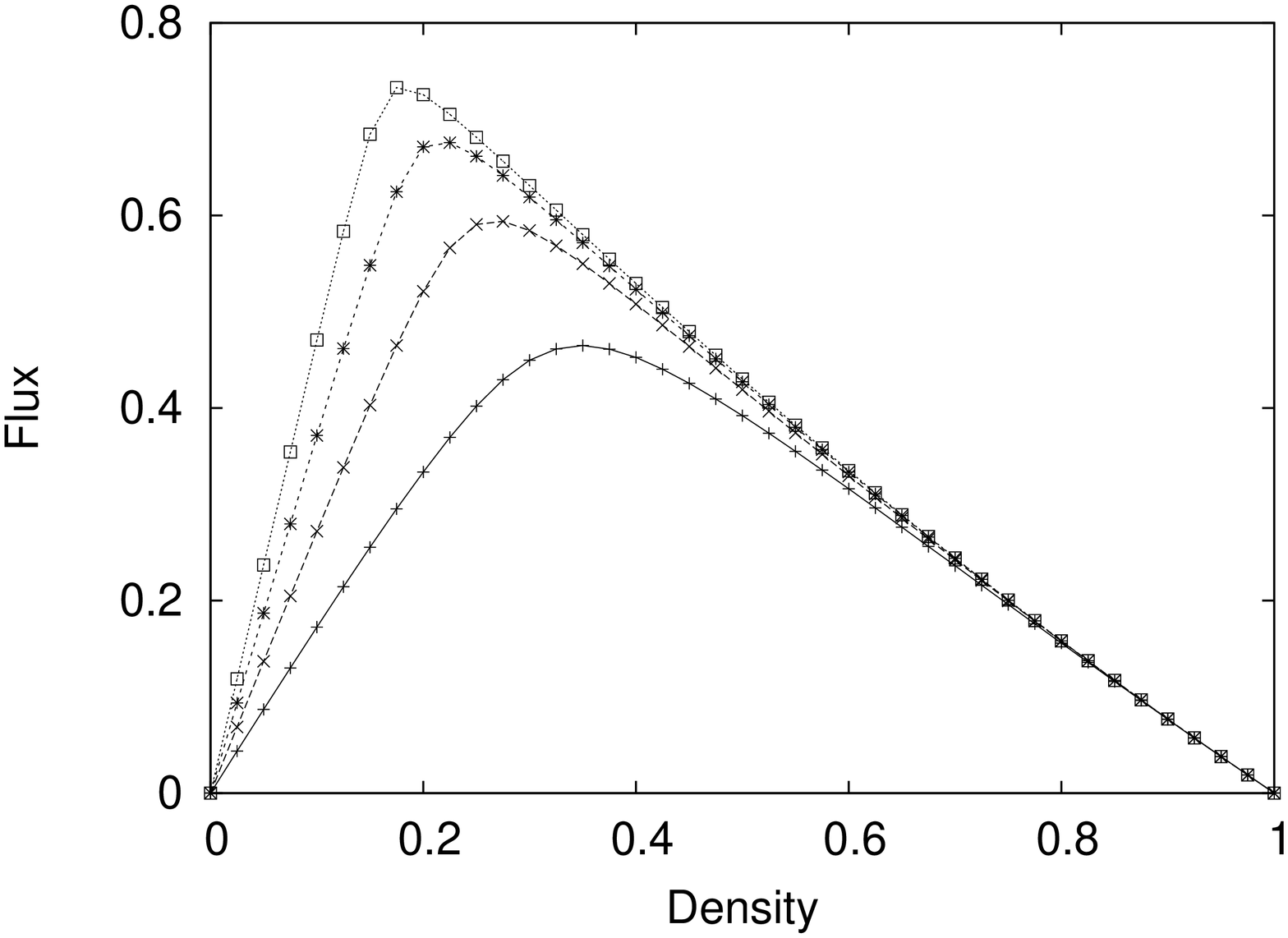}
  \includegraphics[angle=-90,width=0.48\columnwidth]{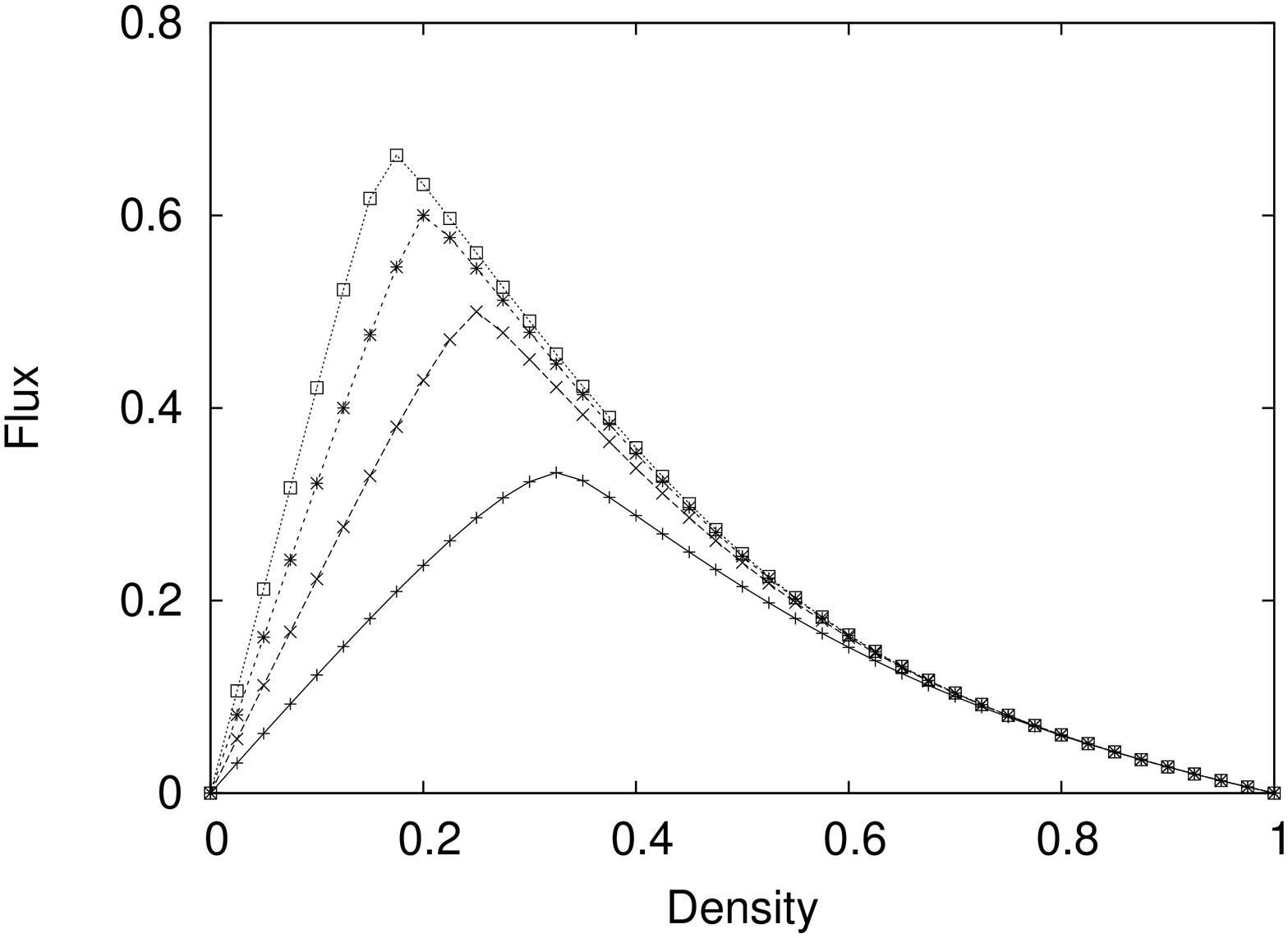}
\end{center}
\hspace{0.5in}(a) \hspace{3.5in}(b)
\caption {Fundamental digram of the ADM for $V_{\rm max} > 1$ corresponding 
to (a) $p=0.25$ and (b) $p=0.75$ respectively,
obtained through computer simulations for 
$V_{\rm max}=2$ ($+$), $V_{\rm max}=3$ ($\times$), 
$V_{\rm max}=4$ ($\ast$), and $V_{\rm max}=5$ ($\Box$), respectively.}
\label{fig-2}
\end{figure}

\begin{figure}[htb]
\begin{center}
\includegraphics[angle=-90,width=0.48\columnwidth]{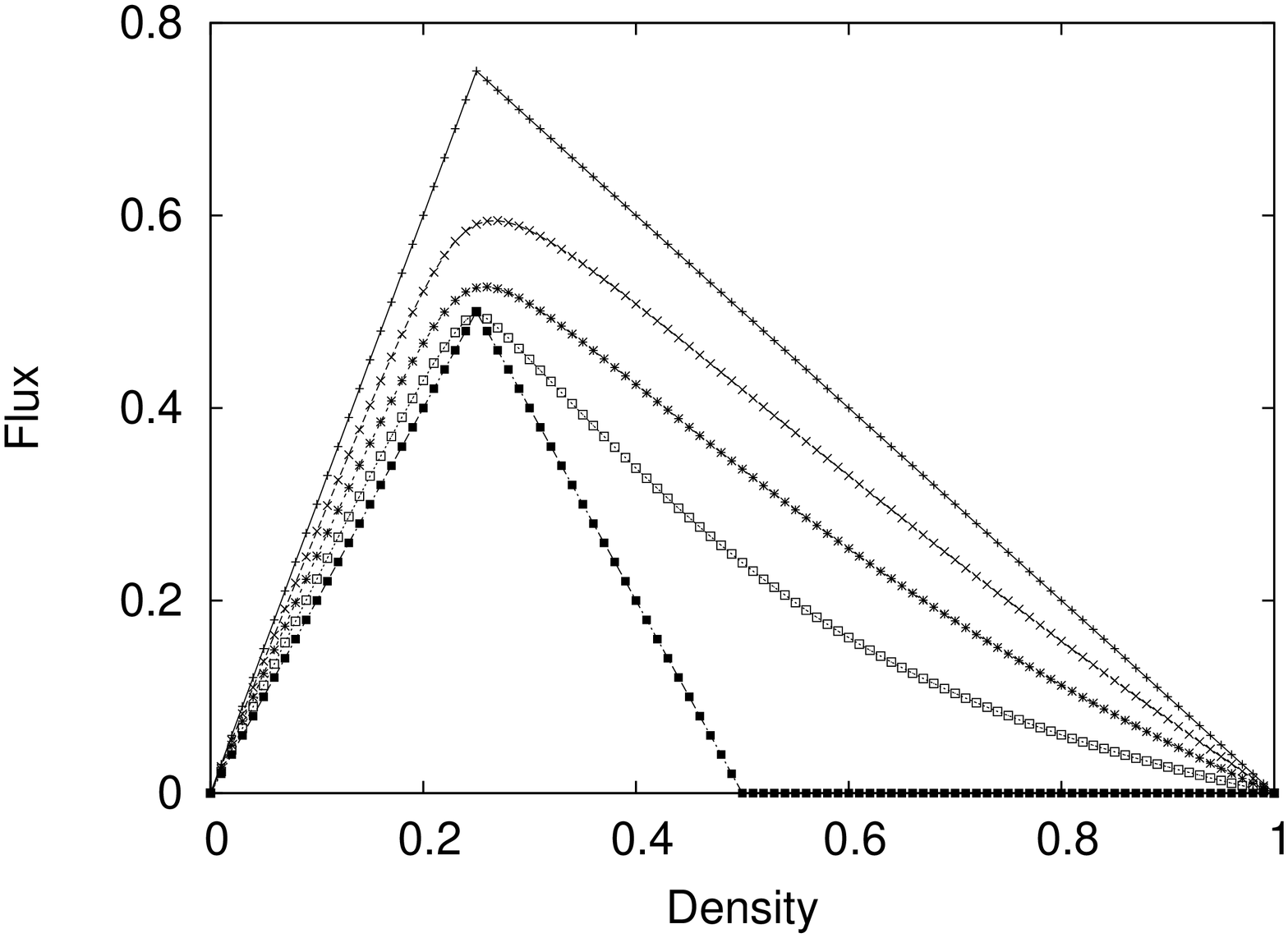}
\includegraphics[angle=-90,width=0.48\columnwidth]{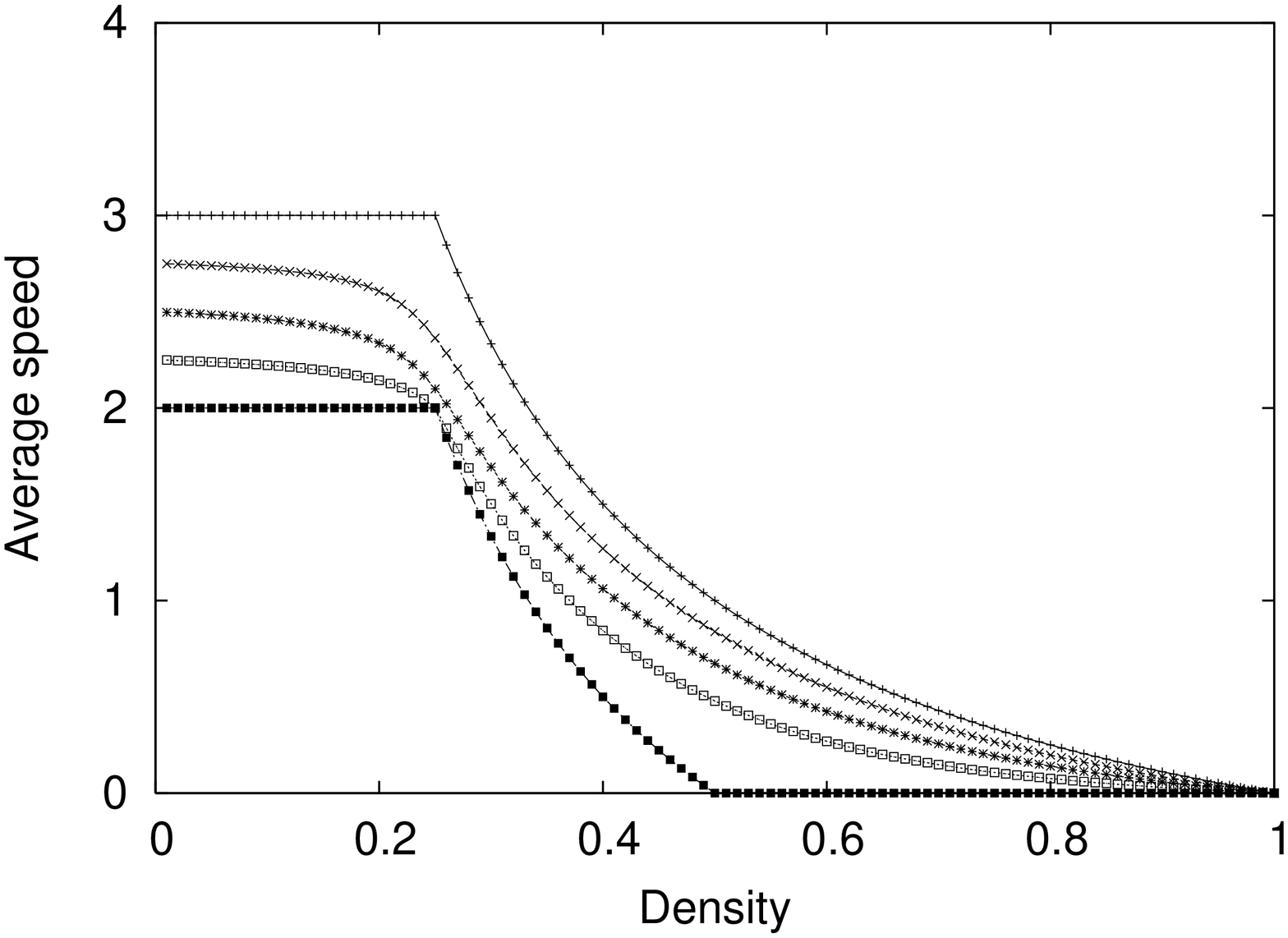}
\end{center}
(a) \hspace{3.5in}(b)
\caption {(a) Fundamental diagram and (b) density-dependence of the
average speed of vehicles of the ADM with $V_{\rm max}=3$ 
for $p=0.0$ ($+$), $p=0.25$ ($\times$), $p=0.50$ ($\ast$), 
$p=0.75$ ($\square$) and $p=1.0$ ($\blacksquare$), respectively.}
\label{fig-3}
\end{figure}

In the special case $V_{\rm max}=1$, the ADM reduces to NaSch model
\cite{nagel} with $V_{\rm max}=1$.  In this limit the fundamental
diagram is given by exact expression \cite{ito1}
\begin{equation}
J= \frac{1}{2} \Big[ 1- \sqrt {1-4(1-p)c(1-c)} \Big ].
\end{equation}
%and is shown in Fig.~\ref{fig-1}. 
The symmetry about $c_{*}=1/2$ in this fundamental diagram breaks down
for all $V_{\rm max} > 1$. Fig.~\ref{fig-2} shows the fundamental diagram
of the ADM for different values of $V_{\rm max}$ for fixed $p=0.25$
and $p=0.75$.  Fig.~\ref{fig-3}(a) and Fig.~\ref{fig-3}(b), show
variation of flux and average speed, respectively, with $c$ for
different values of the braking probability $p$ for fixed $V_{\rm max}=3$.

%%%%%%%%%%%%%%%%%%%%%%%%%%%%%%%%%%%%%%%%%%%%%%%%%%%%%%%%%%%%%%%%%%%
%%%%%%%%%%%%%%%%%%%%%%%%%%%%%%%%%%%%%%%%%%%%%%%%%%%%%%%%%%%%%%%%%%%

%%%%%%%%%%%%%%%%%%%%%%%%%%%%%%%%%%%%%%%%%%%%%%%%%%%%%%%%%%%%%%%%%%%
%%%%%%%%%%%%%%%%%%%%%%%%%%%%%%%%%%%%%%%%%%%%%%%%%%%%%%%%%%%%%%%%%%%

For $V_{\rm max}=1$, the fundamental diagram of ADM has a perfect
particle-hole symmetry with a flow maximum at $c=0.5$.  However, as in
the NaSch model, this particle-hole symmetry breaks down for all $V_{\rm max}
> 1$ and the maximum shifts to lower densities with increasing $V_{\rm
  max}$.  The system remains in the {\it free-flow} regime for
densities on the left side of this maximum where the flux increases
with increasing density.  The densities on the right hand side of 
this maximum correspond to {\it congested flow} regime where flux starts
decreasing with increasing density and finally vanishes at $c=1$. For
a given $V_{\rm max}$, the maximum value of the flux starts decreasing
with increasing braking probability $p$. The fundamental diagram of
the ADM shows unusual behavior in the deterministic limit $p=1$ where
the flux vanishes at $c=0.5$ for all $V_{\rm max} > 1$. The reason for
this unusual behavior will be explained in the following sections.

\begin{figure}[htb]
\begin{center}
\includegraphics[angle=-90,width=0.48\columnwidth]{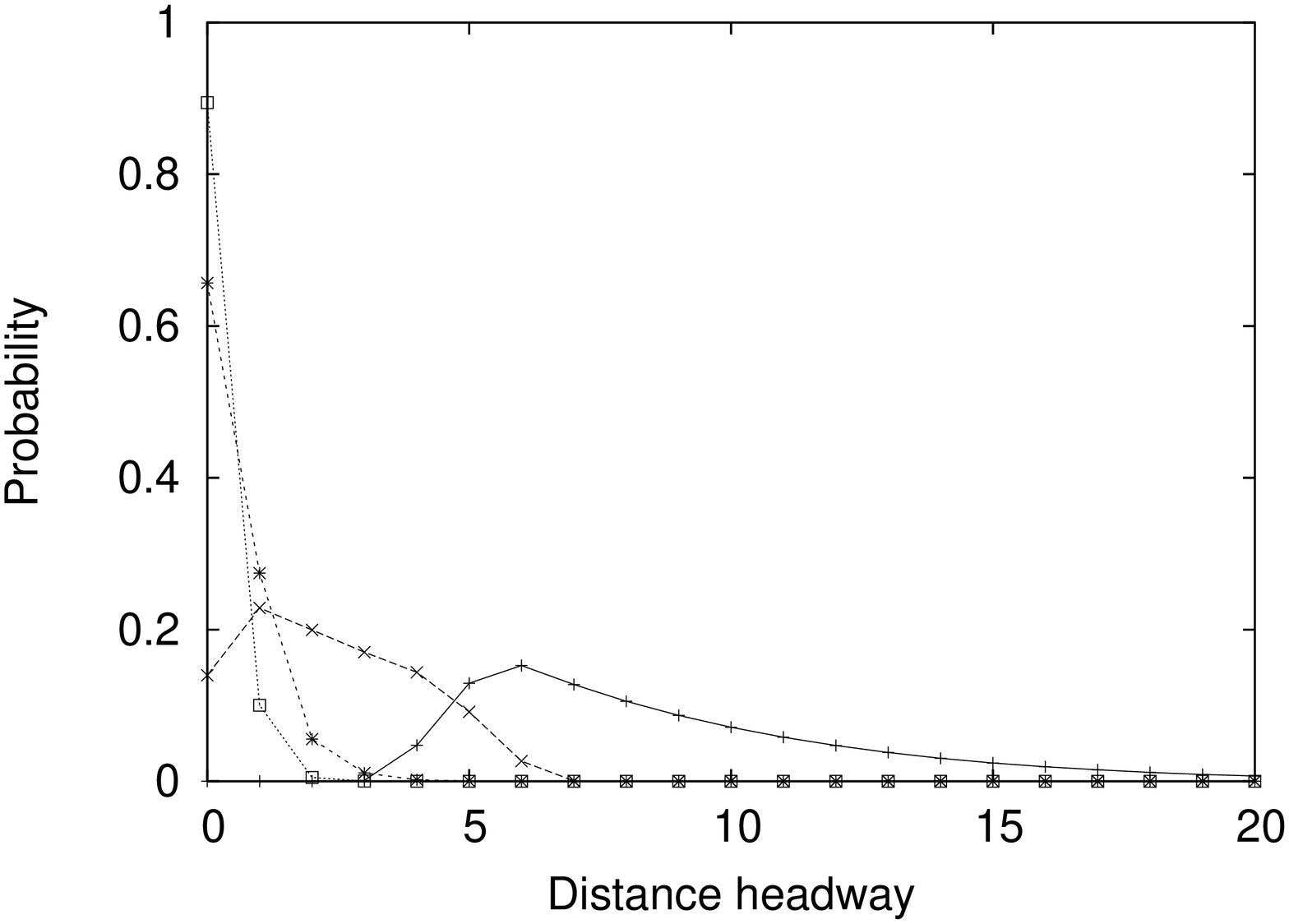}  
\includegraphics[angle=-90,width=0.48\columnwidth]{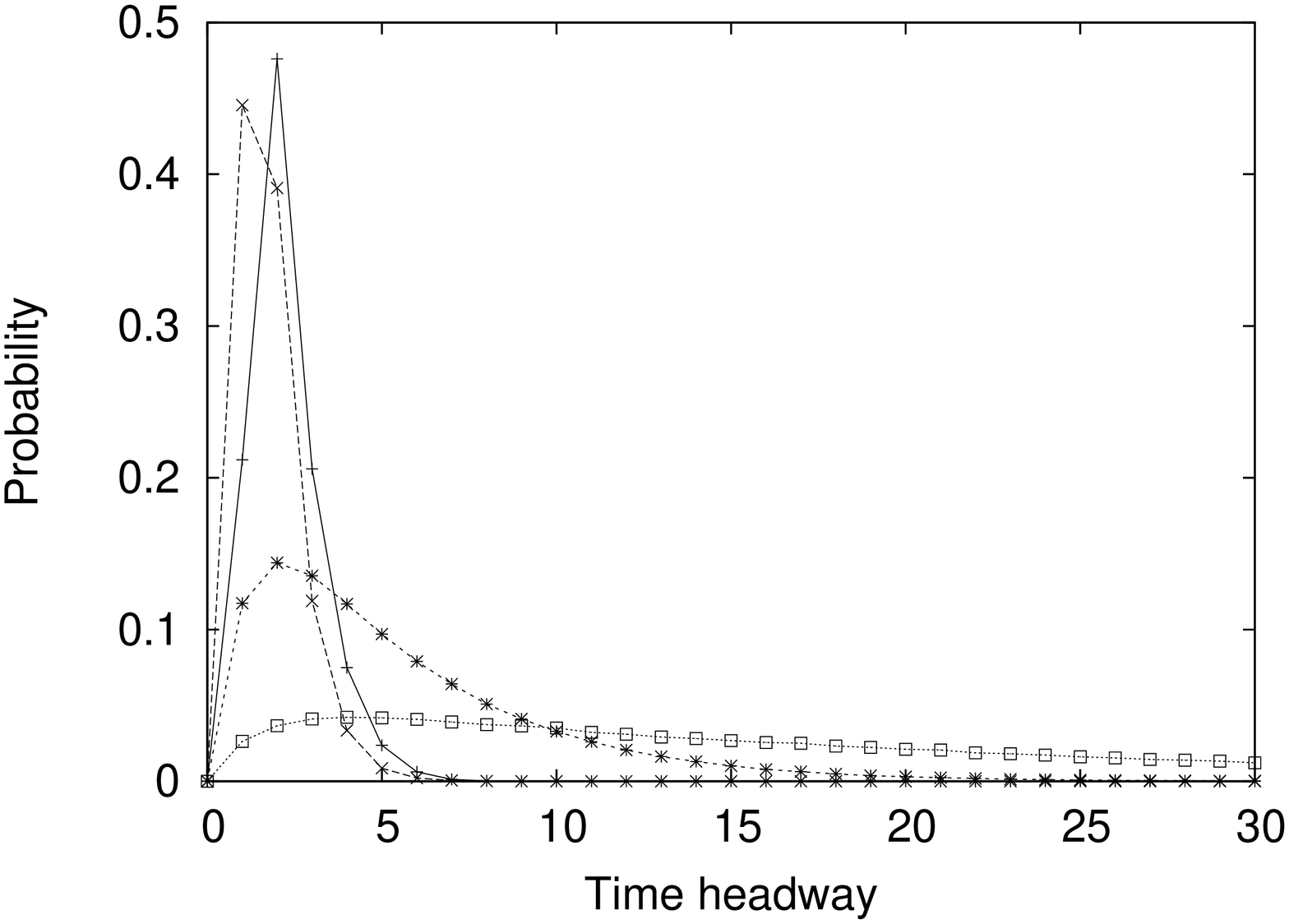}
\end{center}
\hspace{0.5in}(a) \hspace{3.2in}(b)
\caption{Steady-state distributions of (a) distance headways 
and (b) time headways for $p=0.5$ and $V_{\rm max} = 5$
for densities $c=0.1(+)$, $c=0.3(\times)$, $c=0.7(\ast)$,
$c=0.9(\square)$.}
\label{fig-5}
\end{figure}

The {\it distance headway} is usually defined as the distance from a 
selected point on a vehicle to the same point on the corresponding 
lead vehicle (i.e., the next vehicle downstream). Since in our model 
all vehicles have the same length we can use the number $d_n$ of 
empty cells in front of vehicle $n$ as a measure of the headway.
In Fig.~\ref{fig-5}(a) we have shown the distribution $P_{n}$ of the 
distance headway in ADM obtained from simulations. 

At low densities the distance headway distribution shows a broad
peak near $n = V_{\rm max}$. This corresponds to the free-flow 
regime where the cars are distributed almost homogeneously.
In contrast, at higher densities the peak in the distribution 
occurs at a smaller distance headway. In fact, the most probable 
distance headway decreases with increasing density. Finally, at 
sufficiently high densities, the maximum of the probability
distribution occurs only at $n=0$. Thus, with increase of vehicle 
density, the compact cluster of jammed vehicles becomes larger 
while large headways are strongly surpressed.

%%%%%%%%%%%%%%%%%%%%%%%%%%%%%%%%%%%%%%%%%%%%%%%%%%%%%%%%%%%%%%%
%%%%%%%%%%%%%%%%%%%%%%%%%%%%%%%%%%%%%%%%%%%%%%%%%%%%%%%%%%%%%%%

The time headway distribution is determined by (a) the time 
interval between the departure from one site and arrival at the next 
site and (b) the waiting time at a given site; the latter depends 
not only on the hindrance from the vehicle in front but also on the 
randomization parameter $p$. Equivalently, the time-headway depends 
not only on the spatial distance-headway but also on the velocity of 
the vehicles. 

A few typical time headway distributions $P(\tau)$ in ADM are shown 
in Fig.~\ref{fig-5}(b) for a few different densities of the vehicles. 
At sufficiently low densities it shows a peak at $\tau=2$ as, because 
of the {\it parallel} updating scheme, minimum two time steps must 
elapse between the arrival of a vehicle at two successive sites even 
when it moves totally unhindred by any other vehicle. Since mean 
time headway is the inverse of the flux, it is expected to exhibit a 
minimum when plotted against the density. The trend of variation of 
the most probable time headway with increasing density is also similar, 
as can be seen also in Fig.~\ref{fig-5}. At low densities the peak is 
rather sharp and it becomes much broader at higher densities. Compared 
to the corresponding results for the NaSch model \cite{Gosh}, large 
headways are surpressed in the ADM; this is caused by the possibility 
of large acceleration whereas in the NaSch only allowed acceleration 
is unity. The broader distribution at higher densities arises from the 
longer waiting times at each site which is caused by the hindrance from 
the vehicle immediately in front.

%%%%%%%%%%%%%%%%%%%%%%%%%%%%%%%%%%%%%%%%%%%%%%%%%%%%%%%%%%%
\subsection{Numerical and Exact Analytical Results for ADM in Limiting Cases}
%%%%%%%%%%%%%%%%%%%%%%%%%%%%%%%%%%%%%%%%%%%%%%%%%%%%%%%%%%%

%%%%%%%%%%%%%%%%%%%%
\subsubsection{Deterministic limit $p$=0}
%%%%%%%%%%%%%%%%%%%%

This stochastic model becomes deterministic in the limit $p=0$. In this
special case, the deterministic update rules of the model can be written as 
\begin{eqnarray}
V_{n}(t+1) = \min(V_{\rm max},d_{n}) \\
x_{n}(t+1) = x_{n}(t)+ V_{n}(t+1)
\end{eqnarray}
which leads to two types of steady states depending on density of
vehicles \cite{herrmann} . At low densities, the system can
self-organize so that $ d_{n} \ge V_{\rm max}$ for all $n$ and, therefore
every vehicle can move with $V_{\rm max}$, giving rise to the
corresponding flux $cV_{\rm max}$. This steady state is, however,
possible only if enough empty cells are available in front of every
vehicle, i.e., for $c \le c_{*}^{\rm det} = 1/(V_{\rm max}+1)$ and the
corresponding maximum flux is $J_{*}^{\rm det} = V_{\rm max}/(V_{\rm
  max}+1)$.  On the other hand, for $c > c_{*}^{\rm det}$, $d_{n} <
V_{\rm max}$ and, therefore, the relevant steady states are
characterized by $V_{n} = d_{n}$, i.e. flow is limited by density of
holes. Since the average distance headway is $1/c-1$, the fundamental
diagram of the model in the deterministic limit $p=0$ is given by {\it
  exact} expression
\begin{equation}
J = \min[cV_{\rm max},1-c].
\end{equation}
This is identical to the fundamental diagram of the NaSch model
in the deterministic limit, despite the slightly different dynamics.

%%%%%%%%%%%%%%%%%%%%%%%%%%%%%%%%%%%%%%%%%%%%%%%%%%%
\subsubsection{Deterministic limit $p=1$}
%%%%%%%%%%%%%%%%%%%%%%%%%%%%%%%%%%%%%%%%%%%%%%%%%%%

As we discussed earlier in this paper that in the special case $V_{\rm
  max}=1$ the ADM reduces to NaSch model with $V_{\rm max} = 1$ and
hence in the deterministic limit $p=1$, $J=0$ for all densities $c$ as
expected.  However, for $V_{\rm max} > 1$, the properties of the ADM
with maximum allowed speed $V_{\rm max}$ in the deterministic limit
$p=1$ are not exactly identical to those of the same model with
maximum allowed speed $V_{\rm max}-1$ and $p=0$.  If $V_{\rm max} >
1$, then, for $c \ge 1/2$, all initial states lead to $J=0$ because in
the steady state system self-organizes itself in such a way that there
is a maximum headway of one lattice site in front of each vehicle and
hence speed of all vehicles becomes zero immediately after the
randomization step (step $II$ in update rules).  However, for $V_{\rm
  max} > 1 $ and $p=1$, $J \ne 0$ for all $c < 1/2$. The maximal
attainable velocity for every vehicle in this limit becomes $V_{\rm
  max}-1$. The fundamental diagram of ADM for $V_{\rm max} > 1 $ in
the deterministic limit $p=1$ is given by {\it exact} expression \\
\begin{equation}
J=
%\begin{cases}
\cases{
\min[c(V_{\rm max}-1),1-2c] & $\text{\ \ for\ } c \le 1/2$ \\
0 & $\text{\ \ for\ } c \ge 1/2$.
}
%\end{cases}
\end{equation}
This unusual behavior of the ADM is different from the corresponding 
behaviour in the NaSch model. In the deterministic limit $p=1$ of the 
NaSch model, irrespective of $V_{\rm max}$ and $c$, all random initial 
states lead to $J=0$ \cite{debch2}, because a car which has velocity 
$V=0$ will never move again.

%%%%%%%%%%%%%%%%%%%%%%%%%%%%%%%%%%%%%%%%%%%%%%%%%%%%%%%%
\subsubsection{Limit $V_{\rm max} = \infty $}
%%%%%%%%%%%%%%%%%%%%%%%%%%%%%%%%%%%%%%%%%%%%%%%%%%%%%%%%

\begin{figure}[tb]
\begin{center}
\includegraphics[angle=-90,width=0.5\columnwidth]{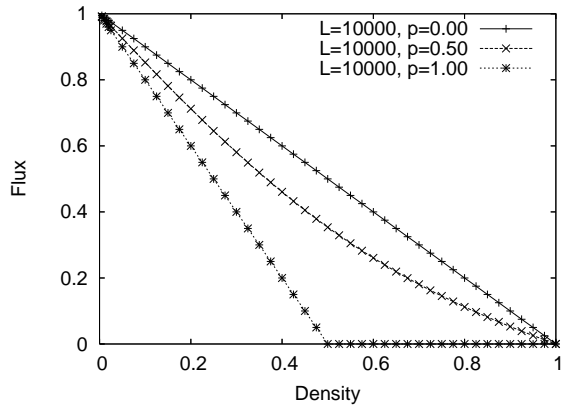}
\end{center}
\caption {Fundamental diagram of the model with $V_{\rm max}=L$ 
for three typical values of $p$. The data only for $L = 10000$ 
have been plotted. The data for $L = 1000$ and $L = 10000$ are 
practically indistinguishable from each other.} 
\label{fig-6}
\end{figure}

There are several possible ways of extrapolating to this limit since only
finite systems can be treated in computer simulations. We here
investigate the case $V_{\rm max}=L$. The fundamental diagram of the
model is plotted in Fig.~\ref{fig-6} for different values of $p$ in
this limit.  This fundamental diagram has a form quite different from
that in the case of finite $V_{\rm max}$. The flow does not vanish in
the limit $c \rightarrow 0$ since already one single vehicle produces
a finite value of flow, $J(c \rightarrow 0) =1 $. $J(c)$ is a
monotonically decreasing function of $c$. Another characteristic
feature of this fundamental diagram is the absence of the characteristic 
plateau which is exhibited by the NaSch model with $V_{\rm max} = \infty$
\cite{sasvari,vinfty}.

%%%%%%%%%%%%%%%%%%%%%%%%%%%%%%%%%%%%%%%%%%%%%%%%%%%%%%%%%%%%%%%%%%
%%%%%%%%%%%%%%%%%%%%%%%%%%%%%%%%%%%%%%%%%%%%%%%%%%%%%%%%%%%%%%%%%%

%%%%%%%%%%%%%%%%%%%%%%%%%%%%%%%%%%%%%%%%%%%%%%%%%%%%%%%%%%%%%%%%%%
\subsection{Approximate analytical theories of ADM}
%%%%%%%%%%%%%%%%%%%%%%%%%%%%%%%%%%%%%%%%%%%%%%%%%%%%%%%%%%%%%%%%%%

In this section we will present the site-oriented mean-field (SOMF)
and car-oriented mean-field (COMF) approaches for calculating the
fundamental diagram of the ADM with periodic boundary conditions 
following the methods of \cite{ito1} after a
brief review of the earlier works done in this regard.  

A SOMF theory was developed earlier for the FI model \cite{bing}.
Starting with a microscopic relation for the updating rule, which
describes the occupancy of each site on the lattice, a macroscopic
time-evolution relation is obtained for the average speed of the
vehicles by carrying out statistical averages.  Mean field equations
are obtained as the asymptotic limit of the evolution relation. This
gives average vehicle speed in the long time limit as a function of
the vehicle density.

A COMF theory for the FI model was developed in
\cite{bing1} starting with the basic equations which describe the time
evolution of the headway in front of each car. By introducing the concept
of inter-car spacing longer and shorter than the maximum attainable
velocity $V_{\rm max}$, the average speed of the vehicles has been
obtained analytically as a function of car density in the asymptotic
limit which corresponds to the steady state.

%%%%%%%%%%%%%%%%%%%%%%%%%%%%%%%%%%%%%%%%%%%%%%%%%%%%%%%%%%%%%%
\subsubsection{Site-oriented Mean-field Theory of ADM}
%%%%%%%%%%%%%%%%%%%%%%%%%%%%%%%%%%%%%%%%%%%%%%%%%%%%%%%%%%%%%%

In the SOMF \cite{ito1} approach, ${c_V}(i,t)$ denotes the probability that
there is a vehicle with speed $V = 0,1,2,...{V_{\rm max}}$ at site
$i$ at time step $t$. Then, obviously, ${c}(i,t) = \sum_{j=0}^{V_{\rm
    max}} {c_j}(i,t)$ is the probability that the site $i$ is occupied
by a vehicle at the time step $t$ (irrespective of its speed) and 
$d(i,t) = 1-c(i,t)$ is the
corresponding probability that the site $i$ is empty. Using the definition \\
\begin{equation}
J(c,p)= \sum_{V=1}^{V_{\rm max}}V {c_V}
\end{equation}
for the flux $J(c,p)$ one can determine the mean-field fundamental
diagram for the given $p$, provided one can determine $c_V$ in the mean-field
approximation.

According to the update rules of the ADM, the time evolutions of the
probabilities ${c_V}(i,t)$ are given by the following equations:\\
{\it Step I:} Acceleration($t \rightarrow {t_1}$)
\begin{eqnarray}
{c_0}(i,{t_1})&=&{c}(i,t){c}(i+1,t) \\ 
%{c_0}(i,{t_1})&=&{c_0}(i,t)+{c_0}(i+1,t) \sum_{V=1}^{V_{\rm max}}{c_v}(i,t)\\ 
{c_V}(i,{t_1})&=&c(i+V+1,t) \prod_{j=1}^{V} d(i+j,t) c(i,t) 
%\sum_{V'=0}^{V_{\rm max}}{c_{v'}}(i,t) \quad
%\hspace{0.5in}
\quad(0 < V < V_{\rm max}) \\
{c_{V_{\rm max}}}(i,{t_1})&=&\prod_{j=1}^{V_{\rm max}} d(i+j,t) c(i,t)
%\sum_{V'=0}^{V_{\rm max}}c_{v'}(i,t)  
\end{eqnarray}

\noindent
{\it Step II:} Randomization (${t_1} \rightarrow {t_2}$)
\begin{eqnarray}
{c_0}(i,{t_2}) &=& {c_0}(i,{t_1}) + p {c_1}(i,{t_1}) \\
{c_V}(i,{t_2}) &=& q{c_V}(i,{t_1}) + p {c_{V+1}}(i,{t_1})\hspace{0.5in}
(0 < V < V_{\rm max})  \\
{c_{V_{\rm max}}}(i,{t_2}) &=& q{c_{V_{\rm max}}}(i,{t_1})
\end{eqnarray} 

\noindent
{\it Step III:} Movement of vehicles (${t_2} \rightarrow t+1$)
\begin{equation}
  {c_{V}}(i,t+1) = {c_V}(i-V,{t_2}) \hspace{0.5in}(0 \le V 
  \le V_{\rm max})
\end{equation}

%%%%%%%%%%%%%%%%%%%%%%%%%%%%%%%%%%%%%%%%%%%%%%%%%%%%%%%%%%%%%%%%%
\begin{figure}[tb]
\begin{center}
\includegraphics[angle=-90,width=0.46\columnwidth]{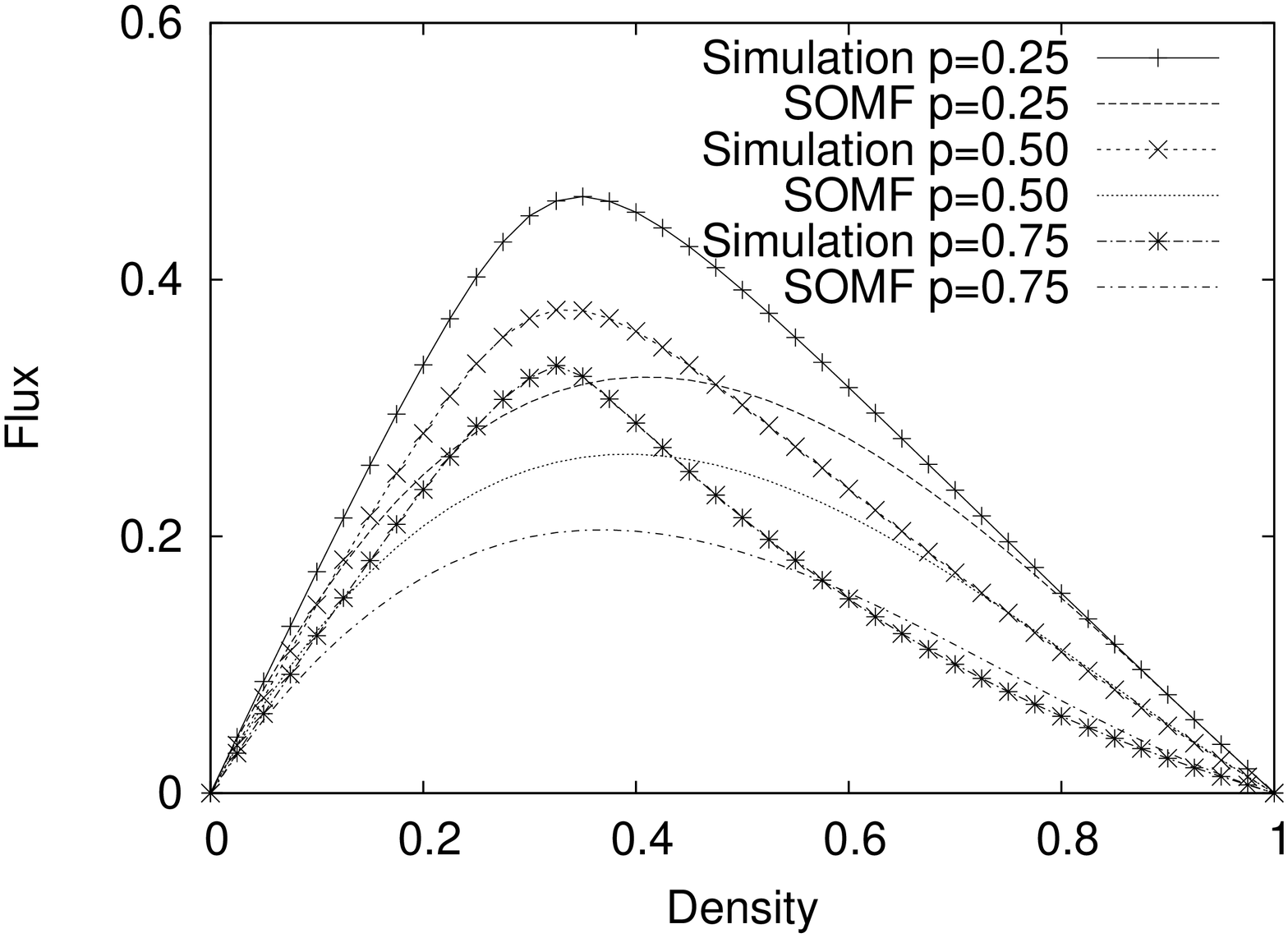}\quad
\includegraphics[angle=-90,width=0.46\columnwidth]{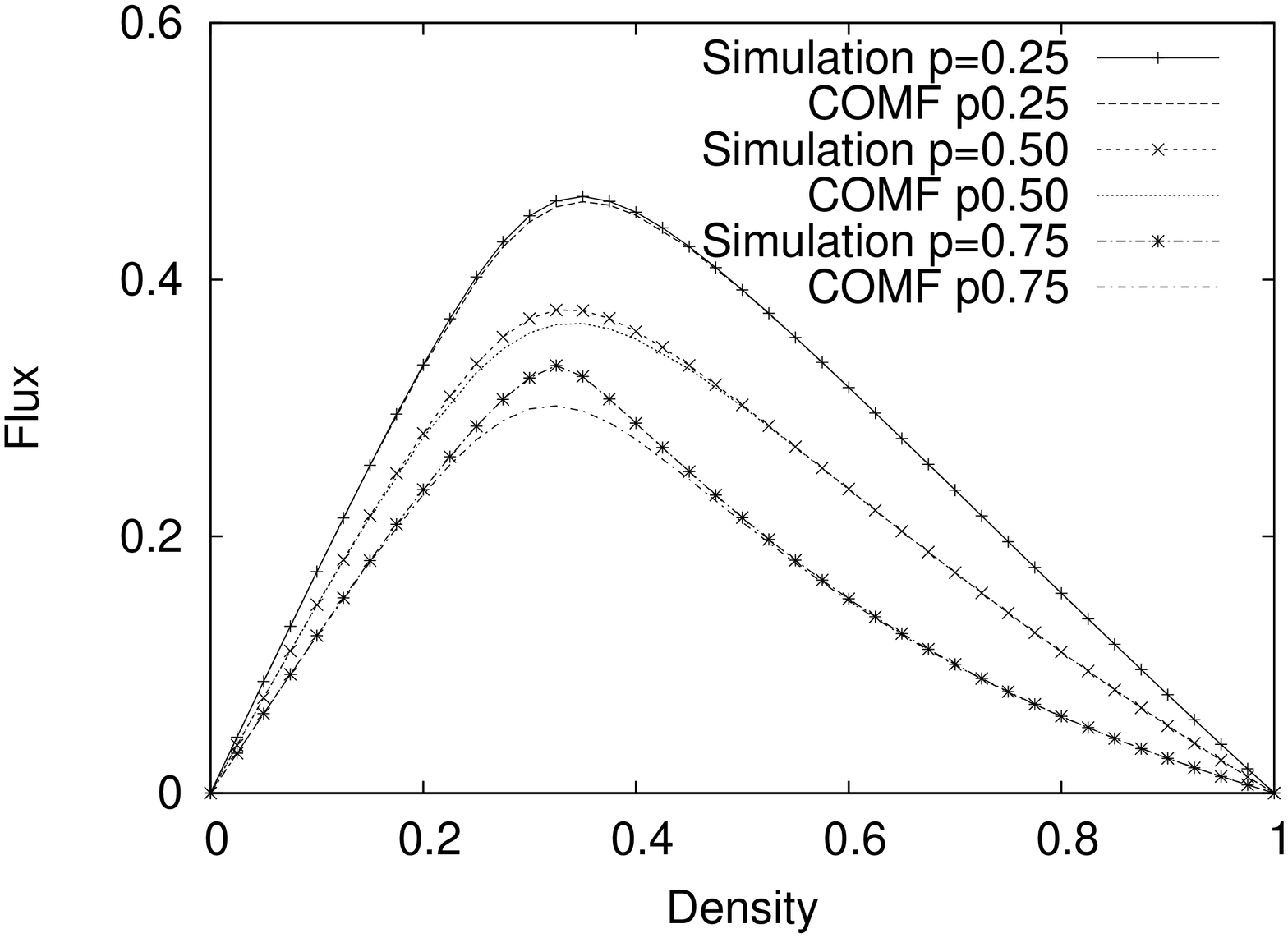}
\end{center}
\caption {Comparison of site-oriented (left) and car-oriented (right)
mean-field theory with results from computer simulation for 
$V_{\rm max} =2$.}
\label{fig-7}
\end{figure}
%%%%%%%%%%%%%%%%%%%%%%%%%%%%%%%%%%%%%%%%%%%%%%%%%%%%%%%%%%%%%%%%%

Recall that ADM with $V_{\rm max} = 1$ is identical to the NaSch 
model with $V_{\rm max} = 1$. Therefore, for nontrivial  results 
of the ADM one must consider $V_{\rm max} \geq 2$.
For $V_{\rm max}=2$, the full SOMF equations read (with $q=1-p$) \\ \\
\begin{eqnarray}
{c_0}(i,t+1) &=& c(i,t)c(i+1,t) + 
pc(i+2,t)d(i+1,t)c(i,t) \label{somf20}\\
{c_1}(i,t+1) &=& qc(i+1,t)d(i,t)c(i-1,t) + pd(i+1,t)d(i,t)c(i-1,t) 
\label{somf21}\\
{c_2}(i,t+1) &=& qd(i,t)d(i-1,t)c(i-2,t) \label{somf22}
\end{eqnarray}

In the steady state, i.e.\ for $t \rightarrow \infty$, the ${c_V}(i,t)$ 
are independent of $t$. For periodic boundary conditions the system becomes
homogeneous in the steady state and hence the $i$-dependence of
${c_V}(i)$ also drops out.
(\ref{somf20})-(\ref{somf22}) then give $c_V$ explicitly
as a function of the density $c$.
The steady state flux for $V_{\rm max}=2$ is then given by \\
\begin{equation}
J = {c_1}+2{c_2} = c(1-c)(2-c-p)
\end{equation}
The results obtained from the SOMF theory are plotted in 
Fig.~\ref{fig-7}(left) for a few values of $p$ along with the
corresponding numerical data from computer simulation.  The agreement 
between the fundamental diagrams obtained from this simple SOMF theory 
and those obtained from computer simulations is quite poor because the 
important correlations between neighboring sites are neglected in this 
approach.

Interestingly, in contrast to the NaSch model \cite{ito1}, the 
fundamental diagram shows an inflection point at intermediate densities. 
This non-convexity of the flow-density relation becomes more pronounced 
for large values of the randomization $p$.

Asymptotically, for large densities $c\approx 1$, the flow in the ADM
will be identical to that in the corresponding NaSch model, i.e.\
$J\approx (1-p)(1-c)$. However, in the NaSch model, and in many other 
traffic models, the flow at {\em any} density $c$ can never exceed 
$(1-p)(1-c)$, the flow on the jammed branch. But, in contrast, because 
of the possibility of large accelaration of the vehicles in the ADM, 
the flow can far excced the value $(1-p)(1-c)$ at intermediate 
densities. Then, for obvious mathematical reasons, any smooth function 
with the asymptotic behaviour $(1-p)(1-c)$ has to exhibit an inflection 
point.

In the NaSch model SOMF always systematically underestimates the true 
flux because of the effective particle-hole attraction \cite{ito1}. 
Surprisingly, the same is not always true in the ADM (see 
e.g.\ $p=0.75$ in Fig.~\ref{fig-7}(left)). 
This indicates that the correlations in the ADM at intermediate densities
are somewhat different from those in the NaSch case. Since now SOMF
{\em overestimates} the flow over a range of density, this indicates 
the presence of effective particle-particle attraction, instead of 
particle-hole attraction in that regime. This is a consequence of the 
large accelerations of the vehicles which lead to a tendency towards 
particle-particle aggregation. This tendency becomes stronger at large 
values of the randomization $p$, where fluctuations that reduce the 
velocity of a car temporarily become more likely.

In the next section we describe an improved mean field theory,
namely car-oriented mean field theory, which takes into account
certain correlations between the sites.

%%%%%%%%%%%%%%%%%%%%%%%%%%%%%%%%%%%%%%%%%%%%%%%%%%%%%%%%%
\subsubsection{Car-oriented Mean-field Theory of ADM}
%%%%%%%%%%%%%%%%%%%%%%%%%%%%%%%%%%%%%%%%%%%%%%%%%%%%%%%%%

Here, we present the car-oriented mean-field (COMF) theory
\cite{schad} of ADM with $V_{\rm max} = 2$. The central quantity 
in COMF theory is the probability ${P_n}(t)$ to find at time $t$ 
(exactly) $n$ empty sites in front of a vehicle, i.e.\ the spatial 
headway distribution. This approach is also known as {\it empty 
interval method} or {\it interparticle distribution function method}. 
For a nice introduction and list of related references we refer to 
\cite{emptyinterval}. 

The time evolution of the probabilities ${P_n}(t)$ can conveniently be
expressed through the probability ${g_j}(t)$ $(j=0,1,2)$ that a car
moves $j$ sites in the next time step.  
In order to find the time evolution of the ${P_n}(t)$ we first
determine from which configurations at time $t$ a given state a time
$t+1$ could have evolved. Take for instance a car --- called second car in
the following --- which has $n \ge 4$ free sites in front, i.e. its
distance to the next car ahead (called first car in the following) is
$n+1$ sites. Since the velocity difference of the two cars is at most 2, a
headway of $n$ sites at time $t+1$ must have evolved from a headway of length
$n-1$, $n$, $n+1$ or $n+2$ in the previous time step. A headway of $n-1$
sites evolves into a headway of $n$ sites only if the first car moves
(with probability ${g_2}(t)$) and the second car brakes in the
randomization step (with probability $p$), i.e. the total probability
for this process is $p{g_2}(t)P_{n-1}(t)$. The headway will remain
constant if the first car moves with probability $g_1(t)$ and second
car brakes with probability $p$ (total probability for this process is
$p{g_1}(t){P_n}(t)$) or the first car moves with probability $g_2(t)$
and second car moves with probability $q$ (total probability for this
process is $q{g_2}(t){P_n}(t)$).  Similarly, a headway of $n+1$ sites
evolves into a headway of $n$ sites if the first car does not move
(probability ${g_0}(t)$)and second car brakes with probability $p$
(total probability being $p{g_0}(t)P_{n+1}(t)$) or the first car moves
with probability $g_1(t)$ and second car moves with probability $q$
(total probability for this process is $q{g_1}(t)P_{n+1}(t)$).
Finally, a headway of $n+2$ evolves into a headway of $n$ only if the second
car moves with probability $q$ (total probability for this
$q{g_0}(t)P_{n+2}(t)$).  

The special cases $n=0,1,2$ and $3$
can be treated in an analogous fashion. In this way one obtains the
time evolution of the probabilities as
\begin{eqnarray}
\label{first}
\fl \qquad{P_0}(t+1) &=& {g_0}(t){P_0}(t) + q{g_0}(t)[{P_1}(t)+{P_2}(t)], \\
\fl \qquad{P_1}(t+1) &=& {g_1}(t){P_0}(t) + (p{g_0}(t)+q{g_1}(t))
[{P_1}(t)+{P_2}(t)] + q{g_0}(t){P_3}(t),\quad \\
\fl \qquad{P_2}(t+1) &=& {g_2}(t){P_0}(t) + (p{g_1}(t)+q{g_2}(t))
[{P_1}(t)+{P_2}(t)] \nonumber \\ 
\fl \qquad &+& (p{g_0}(t)+ q{g_1}(t)){P_3}(t) + q{g_0}{P_4}(t), \\
\fl \qquad {P_3}(t+1) &=& p{g_2}(t)[{P_1}(t)+{P_2}(t)] + 
(p{g_1}(t)+q{g_2}(t)){P_3}(t) \nonumber \\
\fl \qquad &+& (p{g_0}(t)+q{g_1}(t)){P_4}(t) +q{g_0}(t){P_5}(t), \\
\fl \qquad {P_n}(t+1) &=& p{g_2}(t){P_{n-1}}(t) + (p{g_1}(t)+q{g_2}(t))
{P_n}(t) \nonumber \\
\fl \qquad &+& (p{g_0}(t)+q{g_1}(t)){P_{n+1}}(t) + q{g_0}(t){P_{n+2}}(t) \ \ \ 
%\hspace{2.0in} 
(n \ge 4) 
\label{last}
\end{eqnarray}
A car will not move in next time step if there is no empty site in
front of it (probability ${P_0}(t)$) or if there is exactly one empty
site in front of it and it decelerates in the randomization step 2 
(probability $p{P_1}(t)$). It will move one site if either there is
exactly one empty site ahead and it does not decelerate (probability
$q{P_1}(t)$) or there are at least two empty sites in front, but the
car decelerates in step 2 (probability $p\sum_{n\ge2}{P_n}(t)$). In
all other cases it will move two sites.
Therefore the probability $g_j(t)$ that a car moves $j$ sites in the
next time step is given by
\begin{eqnarray}
{g_0}(t) &=& {P_0}(t) + p{P_1}(t) \nonumber \\
{g_1}(t) &=& q{P_1}(t) + p\sum_{n\ge2}{P_n}(t) = p-p{P_0}(t)+(q-p){P_1}(t) \\
{g_2}(t) &=& q\sum_{n\ge2}{P_n}(t)= q[1-{P_0}(t)-{P_1}(t)] \nonumber 
\end{eqnarray}
where we have used the normalization condition 
\begin{equation}
\label{second}
\sum_{n\ge0}{P_n}(t) = 1
\end{equation}
to rewrite the probabilities $g_j(t)$ in terms of ${P_0}(t)$, ${P_1}(t)$, 
${P_2}(t)$ and  ${P_3}(t)$ only.

The probabilities can also be related to the density $c=N/L$ of cars.
Since each car which has the headway $n$ to the next car one in front
of it 'occupies' $n+1$ sites we have following relation:
\begin{equation}
\sum_{n\ge0}(n+1){P_n}(t) = \frac{1}{c}\, .
\end{equation}
Here we are mainly interested in the stationary state (t $\rightarrow
\infty$) with $\lim_{t \rightarrow \infty}$ ${P_n}(t) = {P_n}$. In
order to determine the probabilities in the stationary state we
introduce the generating function \\
\begin{equation}
 P(z) = \sum_{n=0}^{\infty}{P_n}z^{n+1}
\end{equation}
After multiplying corresponding equation in [\ref{first}-\ref{last}]
by $z^{n+1}$ and summing over all equations one finds an explicit
expression for the generating function,
\begin{equation}
\label{generating}
P(z)=\frac{{a_5}{z^5}+{a_4}{z^4}+{a_3}{z^3}+{a_2}{z^2}+{a_1}z}{-pg_2({z^2}
-{2b_1}{z}+{b_2})}
\end{equation}
with 
%\begin{eqnarray}
%a_5 &=& pg_2P_1 \nonumber \\
%%
%a_4 &=& g_2P_0 + (pg_1+qg_2)P_1 \nonumber \\ 
%%
%a_3 &=& (g_1+qg_2)P_0 + (pg_0+qg_1)P_1 \\
%%
%a_2 &=& (g_0+qg_1)P_0+qg_0P_1 \nonumber \\
%%
%a_1 &=& qg_0P_0 \nonumber 
%\end{eqnarray}
%\begin{alignat}{2}
\begin{eqnarray}
\fl a_1 = qg_0P_0, \qquad && a_2 = (g_0+qg_1)P_0+qg_0P_1 ,\nonumber\\
\fl a_3 = (g_1+qg_2)P_0 + (pg_0+qg_1)P_1, \qquad
&& a_4 = g_2P_0 + (pg_1+qg_2)P_1,  \qquad a_5 = pg_2P_1,  \nonumber\\
\fl b_1 = \frac{qg_1+g_0}{2pg_2}, \qquad 
&& b_2 = -\frac{qg_0}{pg_2}.
%\nonumber
%\end{alignat}
\end{eqnarray}
Note that $\sum_{j}a_{j} = (1+q){P_0} + {P_1}$.
%and 
%\begin{eqnarray}
%b_1 = \frac{(qg_1+g_0)}{2pg_2}\hspace{0.5in} b_2 = -\frac{qg_0}{pg_2} .
%\end{eqnarray}
The denominator of $P(z)$ has two zeros located at $s_{\pm}=b_1 \pm
\sqrt{b_{1}^{2}-b_2}$ with $\vert s_{+} \vert \ge 1$ and $\vert s_{-}
\vert \le 1$. 

The normalization condition (\ref{second}) is equivalent to $P(1)=1$
and is already satisfied by (\ref{generating}). The density relation
implies $P'(1)=\frac{1}{c}$, where $P'(z)$ denotes the derivative of
$P(z)$. In order to have $0 \le P_n \le 1$ and $\lim_{n \rightarrow
  \infty} P_n =0 $ the generating function must be analytic in the
unit disc $\vert z \vert \le 1$. Therefore the zero $s_{-}$ of the
denominator has to be cancelled by a corresponding zero of the
numerator. The equation
${a_5}{s_{-}^5}+{a_4}{s_{-}^4}+{a_3}{s_{-}^3}+{a_2}{s_{-}^2}+{a_1}s_{-}=0$
yields a relation between the variable $P_0$ and $P_1$ so that $P(z)$
only depends on one free parameter, e.g. $P_0$. This parameter, in
turn, is a function of the only physically relevant parameter, the
density c, via $P'(1)=\frac{1}{c}$.

To obtain the fundamental diagram we have to calculate the flux. It is
given by 
\begin{equation}
J(c,p) = c[g_1 + 2g_2].
\label{eq-29}
\end{equation}

In order to calculate the flux $J(c,p)$ for a given set of $c$ and $p$ one
has to solve the following two equations numerically. 
\begin{equation}
\label{appenda1}
{a_5}{s_{-}^5}+{a_4}{s_{-}^4}+{a_3}{s_{-}^3}+{a_2}{s_{-}^2}+{a_1}s_{-}=0
\end{equation}
and 
\begin{equation}
\label{pdashone}
P'(1)=\frac{1}{c}
\end{equation}
where $P'(z)$ denotes the derivative of $P(z)$.
\noindent
Eq.~(\ref{pdashone}) can be written as
\begin{equation}
\label{appenda3}
\frac{p{g_2}(2-2{b_1})+(5{a_5}+4{a_4}+3{a_3}+2{a_2}+{a_1})}{p{g_2}(1-2{b_1}
+{b_2})}+\frac{1}{c}=0
\end{equation}
Eqs.~(\ref{appenda1}) and (\ref{appenda3}) were solved numerically.
Values of $P_0$ and $P_1$ thus obtained for a given set of $c$ and $p$ 
are used to calculate the values of $g_0$, $g_1$ and $g_2$. Finally 
flux $J$ is calculated using equation (\ref{eq-29}).

The results obtained from COMF are plotted in Fig.~\ref{fig-7}(right)
for a few values of $p$ along with the corresponding numerical data
from computer simulation.  Fundamental diagrams obtained from COMF
show an excellent agreement with the numerical data in the limit $p
\rightarrow 0$. 
Thus COMF can capture the important correlations much better
than SOMF. Especially it is able to reproduce the occurance of an
inflection point at larger values of $p$.
The small deviations are due to the fact that COMF
neglects the correlations between the headways in front of successive
vehicles.

%%%%%%%%%%%%%%%%%%%%%%%%%%%%%%%%%%%%%%%%%%%%%%%%%%%%%%%%%%%%%%%%%%
\subsection{Comparison of ADM with Nagel-Schreckenberg Model}
%%%%%%%%%%%%%%%%%%%%%%%%%%%%%%%%%%%%%%%%%%%%%%%%%%%%%%%%%%%%%%%%%

%%%%%%%%%%%%%%%%%%%%%%%%%%%%%%%%%%%%%%%%%%%%%%%%%%%%%%%%%%%%%%%%%
\begin{figure}[htb]
\begin{center}
\includegraphics[angle=-90,width=0.49\columnwidth]{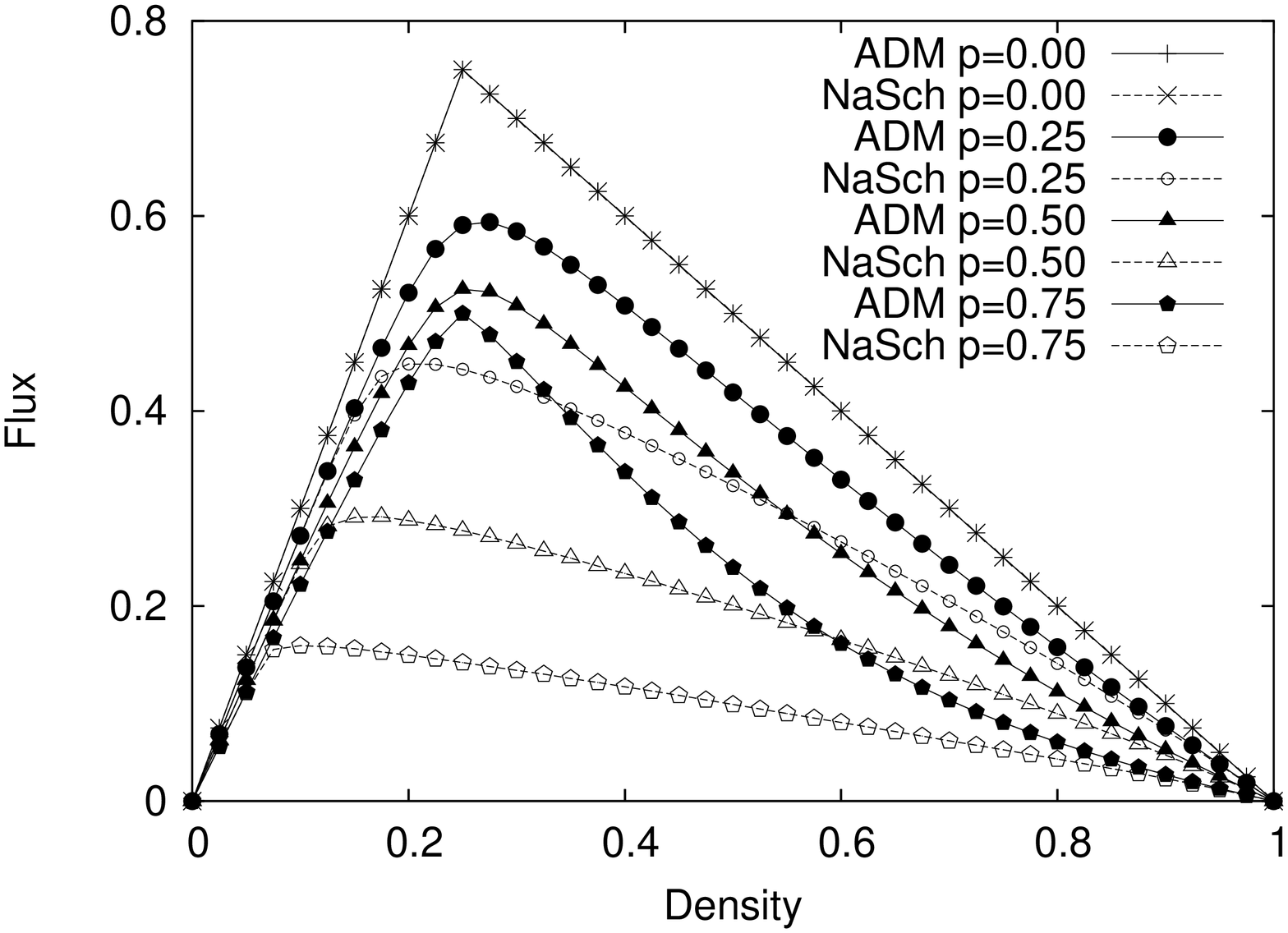}\quad
\includegraphics[angle=-90,width=0.49\columnwidth]{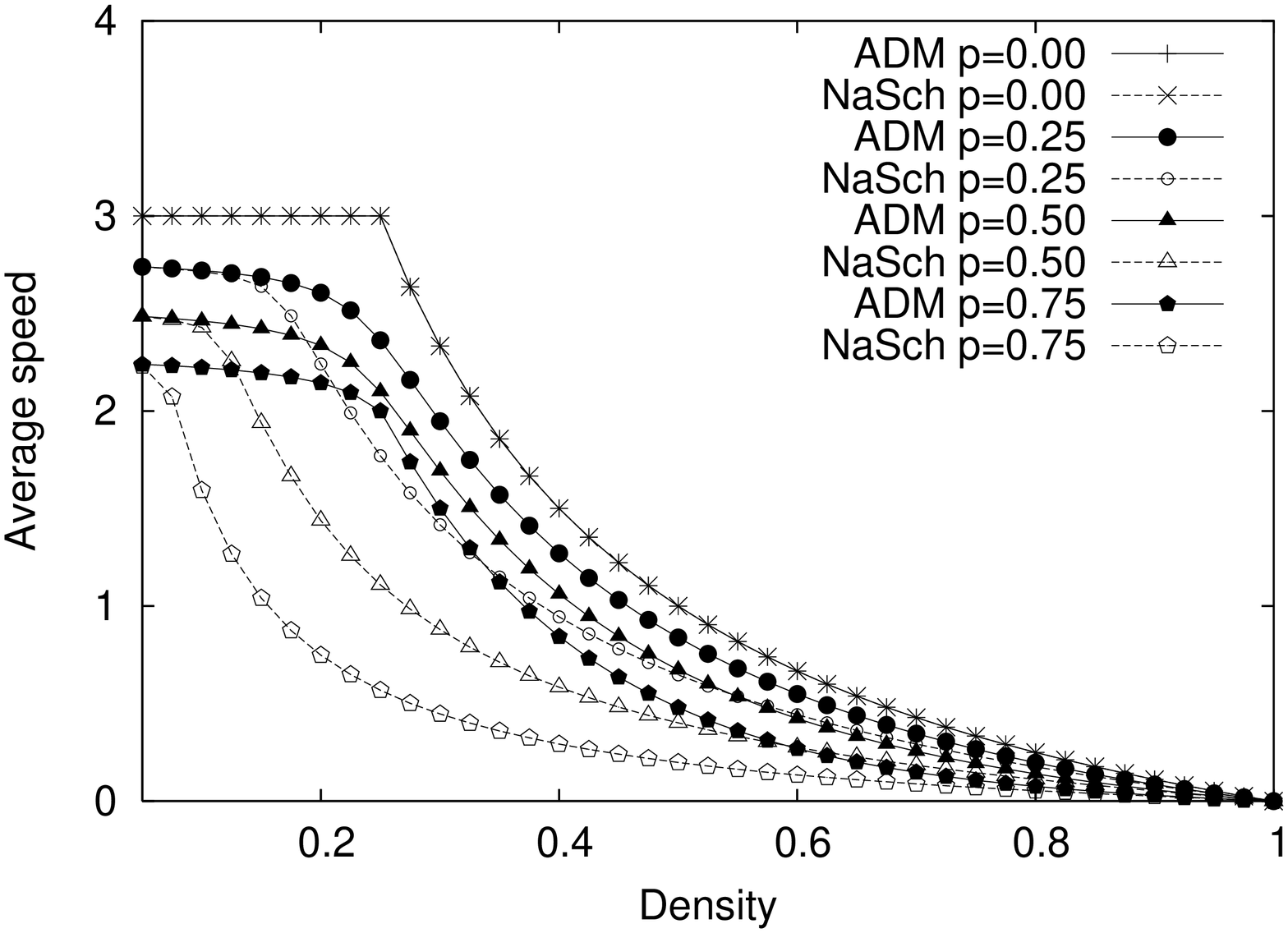}
\end{center}
\phantom{asasasda}(a) \hspace{3.5in}(b)
\caption {(a) The fundamental diagram of the model plotted along with the
  fundamental diagram of NaSch model for $V_{\rm max} =3$ for a few
  values of $p$.  (b) Average speed of vehicles plotted against their
  densities for the model of aggressive driving along with the NaSch
  model for $V_{\rm max}=3$ for a few values of $p$. 
}
\label{fig-9}
\end{figure}
%%%%%%%%%%%%%%%%%%%%%%%%%%%%%%%%%%%%%%%%%%%%%%%%%%%%%%%%%%%%%%%%%

In Fig.~\ref{fig-9} we have plotted the fundamental diagram of the ADM
with $V_{\rm max} =3$ and the average speed of vehicles against their
density along with that of the NaSch model with $V_{\rm max} =3$ for a
few values of $p$. In the absence of randomization, i.e. for $p=0$,
this model and the NaSch model give identical fundamental diagram and
variation of average speed with density. In the presence of
randomization, i.e. $p \ne 0$, the flow in the ADM is always
larger than that of the corresponding NaSch model due to the
faster acceleration. This difference is most pronounced at densities
slighty beyond the maximum flow.

%%%%%%%%%%%%%%%%%%%%%%%%%%%%%%%%%%%%%%%%%%%%%%%%%%%%%%%%%%%%%%%%%%%%%%%%%%%%%%
\section{ADM with Open Boundary Conditions} 
%%%%%%%%%%%%%%%%%%%%%%%%%%%%%%%%%%%%%%%%%%%%%%%%%%%%%%%%%%%%%%%%%%%%%%%%%%%%%%

%%%%%%%%%%%%%%%%%%%%%%%%%%%%%%%%%%%%%%%%%%%%%%%%%%%%%%%%%%%%%%%%%%%%%%%%%%%%%%
\label{sec4}
\begin{figure}[htb]
\begin{center}
\includegraphics[angle=0,width=0.75\columnwidth]{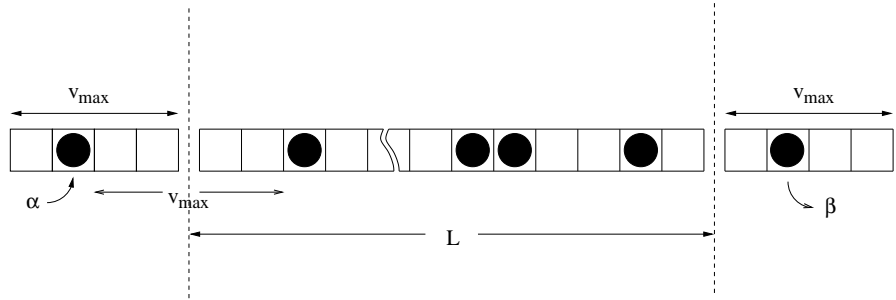}
\end{center}
\caption {Schematic representation of the analyzed system with
  open boundary conditions. The main system consists of $L$ cells. 
  Vehicles move from left to right, and are represented by dark 
  circles. The left boundary consists of mini system of 
  $V_{\rm max}$ cells. This left mini system is occupied by
  at most one vehicle with probability $\alpha$. Similarly the right
  boundary consists of a mini system of $V_{\rm max}$ cells and
  particles are extracted from it with probability $\beta$. We 
  shall represent each cell by a site on the lattice formed by 
  these cells.}
\label{fig-10}
\end{figure}
%%%%%%%%%%%%%%%%%%%%%%%%%%%%%%%%%%%%%%%%%%%%%%%%%%%%%%%%%%%%%%%%%%%%%%%%%%%%%%

In this section we consider the ADM with open boundary conditions
where vehicles move deterministically, i.e.\ with randomization probability
$p=0$. 
%Results for the case with non-zero randomization
%probability ($p>0$) will be published elsewhere \cite{kunwar}. 

A schematic representation of the analyzed system is shown in
Fig.~\ref{fig-10}.  Our main system consist of $L$ cells. This main
system is connected to two mini systems of length $V_{\rm max}$ on
each side \cite{huisinga}.  This is done to provide a proper insertion
and extraction strategy allowing us to investigate the whole spectrum
of the possible states.  The state of the mini system of the left
boundary has to be updated every time step before the vehicles of
whole system. The update procedure consists of two steps. If any cell
of the left mini system is occupied it has to be emptied first. Then a
vehicle is inserted in the system with probability $\alpha$. The
position of the inserted vehicle has to satisfy the following
conditions: (i) The headway between the inserted vehicles in the mini
system and the first vehicle in the main system is equal to $V_{\rm
  max}$, and (ii) the distance to the main system has to be minimum
i.e. if there is no vehicle present in the main system within first
$V_{\rm max}$ cell then the rightmost cell of the left boundary is
occupied.  The right boundary consists of $V_{\rm max}$ cells and
vehicle are removed from these cells with probability $\beta$. 
These boundary conditions are capable of generating all flows
observed in the case of periodic boundary conditions, including the
maximal flow. From now onwards, we shall represent the cells by the
sites of a lattice formed by the cells.

The above insertion and extraction scheme generates the maximum flow
of the corresponding aggressive driving model with periodic boundary
conditions for $\alpha = \beta = 1$, i.e.
\begin{equation}
J = \frac{V_{\rm max}}{V_{\rm max}+1}\hspace{0.9in}({\rm for}\hspace{0.1in} 
\alpha = \beta =1).
\end{equation}

%%%%%%%%%%%%%%%%%%%%%%%%%%%%%%%%%%%%%%%%%%%%%%%%%%%%%%%%%%%
\subsection{Density Profiles in the ADM}
%%%%%%%%%%%%%%%%%%%%%%%%%%%%%%%%%%%%%%%%%%%%%%%%%%%%%%%%%%%

\begin{figure}[htb]
\begin{center}
\includegraphics[angle=-90,width=0.48\columnwidth]{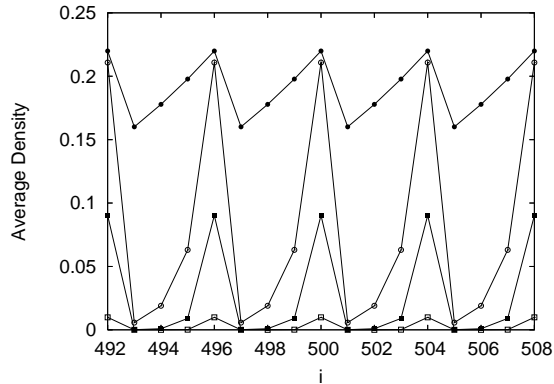}
\end{center}
\caption {Density profile in the middle of the bulk of the system;
$V_{\rm max}=4$ and $L=1000$. The different curves correspond to 
different sets of values of $\alpha$ and $\beta$, namely, 
$\alpha$=0.01 and $\beta$=1.0 ($\square$), $\alpha$=0.1 and $\beta$=1.0 
($\blacksquare$), $\alpha$=0.3 and $\beta$=1.0 ($\circ$), 
$\alpha$=0.9 $\beta$=1.0 ($\bullet$).}
\label{fig-12}
\end{figure}
For small $\alpha$ and large $\beta$, the system is found in the 
free flow regime. In Fig.~\ref{fig-12} we have shown density 
profiles over a spatial region located in the middle of the bulk 
of the system for $V_{\rm max}=4$ and $L=1000$ in the free flow 
regime. The density profile shows a periodic structure with a 
period of oscillation $\Delta i=4$. For any arbitrary $V_{\rm max}$, 
we find that the period of this oscillating pattern is 
$\Delta i= V_{\rm max}$. 

%%%%%%%%%%%%%%%%%%%%%%%%%%%%%%%%%%%%%%%%%%%%%%%%%%%%%%%%%%%%%
\begin{figure}[htb]
\begin{center}
\includegraphics[angle=-90,width=0.48\columnwidth]{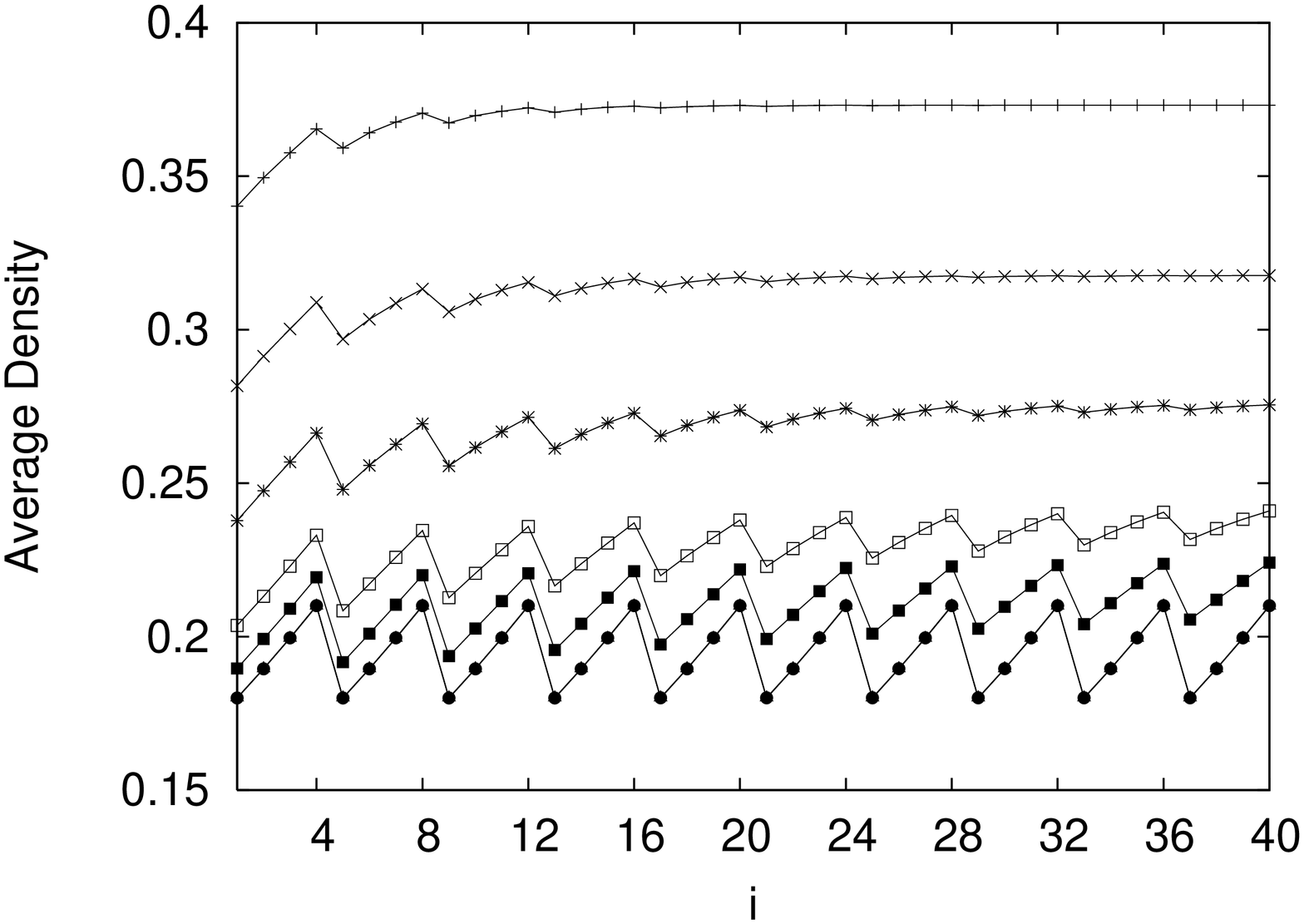}
\includegraphics[angle=-90,width=0.48\columnwidth]{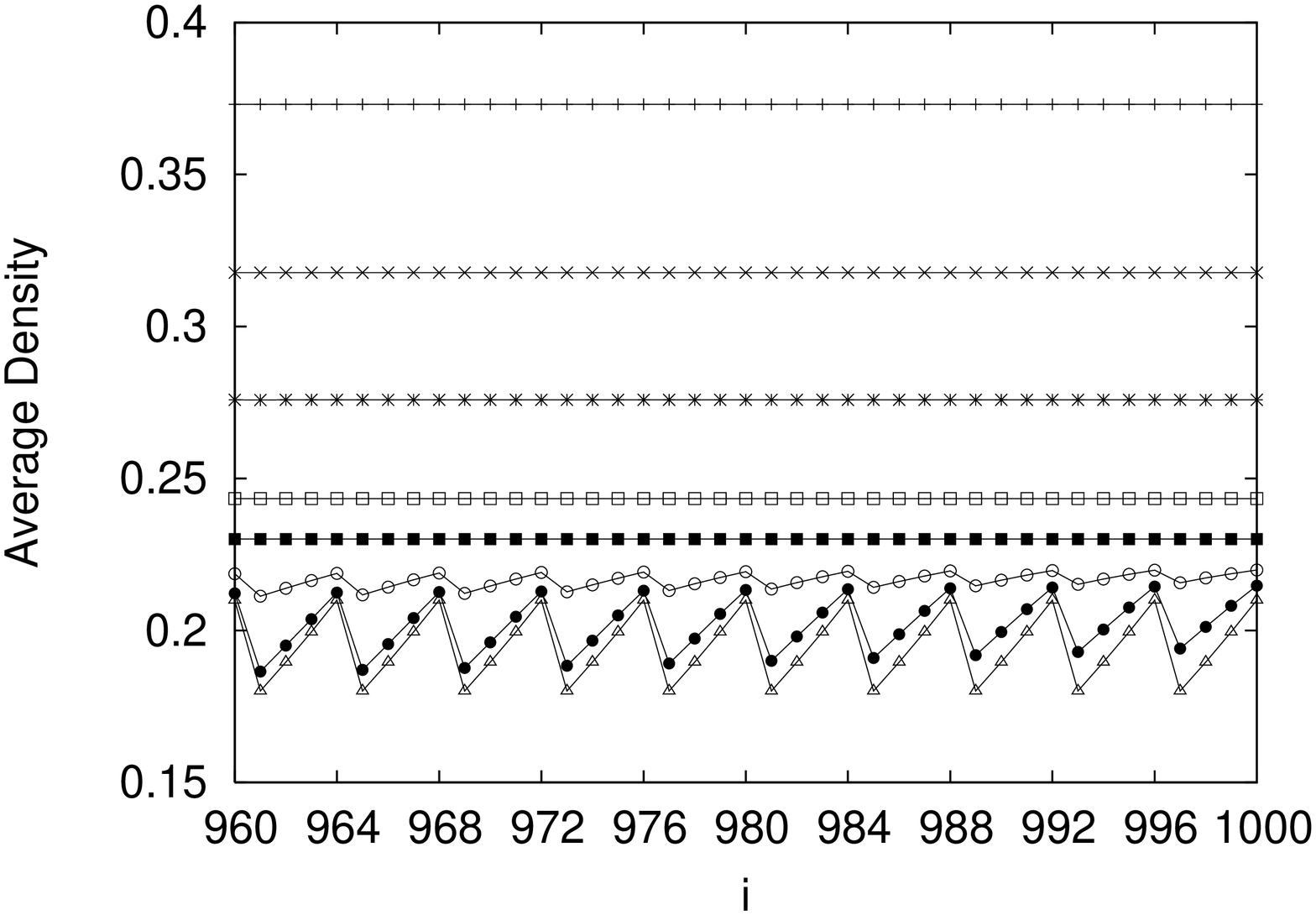}
\end{center}
\phantom{asasasda}(a) \hspace{3.5in}(b)
\caption {(a) Density profile at the beginning of the system
(b) Density profile at the end of the system for $V_{\rm max}=4$ and $L=1000$. 
$\alpha$=0.95 and $\beta$=0.50
($+$), $\alpha$=0.95 and $\beta$=0.60 ($\times$), $\alpha$=0.95 and
$\beta$=0.70 ($\ast$), $\alpha$=0.95 and $\beta$=0.80 ($\square$),
$\alpha$=0.95 and $\beta$=0.85 ($\blacksquare$), $\alpha$=0.95 and
$\beta$=0.90 ($\circ$), $\alpha$=0.95 and $\beta$=0.95 ($\bullet$),
$\alpha$=0.95 and $\beta$=1.00 ($\triangle$).}
\label{fig-13}
\end{figure}
%%%%%%%%%%%%%%%%%%%%%%%%%%%%%%%%%%%%%%%%%%%%%%%%%%%%%%%%%%%%%

In order to understand 
this periodicity we first consider the density profile for very low
injection rates and maximum extraction rate ($\beta=1$) (see
Fig.~\ref{fig-12}).
For $\alpha=0.01$ the probability of inserting a vehicle at the
rightmost site of the left mini-system in two successive time step is
very small and, therefore, the vehicles at the beginning of the system
do not feel influence of each other.  This means that a vehicle which
is inserted at the rightmost site of the left mini system moves to
$i=4$ at the next time step and can be found at site $i=4t$ after $t$
times steps ($t=1,2,3,...$). The density on these sites is $\rho
\approx \alpha$ ($\alpha \le 0.1$).  

For increasing injection
rate $\alpha$, the probability of inserting a vehicle in two
successive time steps increases which results in the increase in the
hindrance that a vehicle feels from the front vehicle at the beginning
of the system. This can be explained as follows \cite{cheybanipre7}: 
\newline 
Suppose we
insert a vehicle A in the mini-system at time step $T$ and a vehicle
at time step $T+1$. Considering the system at time step $T+1$, we see
that vehicle A is on site $i=4$ and will move with velocity $4$ where
as vehicle B will occupy position $i=3$ in the next time step $T+2$
because the vehicles are inserted in left mini-system in such a way
that the headway between the inserted vehicle and the next vehicle
downstream is is $4$. At time $T+t$, vehicle A is on $i=4t$ and
vehicle B is on $i=4(t-1)-1$. In other words, we can say that the
hindrance due to left boundary condition leads to a shift of the
position of the vehicles within the system. This shift is reflected in
the oscillation in the density profile of \ref{fig-12}. The
probability of finding a vehicle at $i=4t+3$ is smaller than at $i=4t$
and it is much smaller for $i=4t+2$ and even much smaller for
$i=4t+1$.

As we move from the free flowing regime to congested flow regime
(keeping $\alpha$ fixed and decreasing $\beta$) something interesting
happens: the oscillations start vanishing and envelope of density
profile rises (see Fig.~\ref{fig-13}).  For low values of $\beta$, the
density profile is just a constant whose value increases with
decreasing $\beta$. This phenomenon is due to the hindrance that
vehicles feel at the right boundary with decreasing probability
$\beta$. Consequently a jam develops at the right boundary which
expands to the left with decreasing $\beta$. 
% (as shown in Fig.  \ref{fig-14}).

%%%%%%%%%%%%%%%%%%%%%%%%%%%%%%%%%%%%%%%%%%%%%%%%%
\subsection{Phase Diagram in the ADM}
%%%%%%%%%%%%%%%%%%%%%%%%%%%%%%%%%%%%%%%%%%%%%%%%%

In order to identify the regions of free-flow and congested flow in
the phase diagram of the ADM we measure the bulk density and the flux
in the middle of the open system by varying the boundary rates. 
Density-flux pairs falling on the free flow branch of the periodic
system are identified as belonging to the free flow phase, those falling
on the jammed branch as congested flow. Since density profile shows 
a periodic structure in the free flow regime, in order to compute the 
bulk density in the middle of the system, we average over the densities 
of $V_{\rm max}$ lattice sites (i.e. one period of oscillation) for a 
given $V_{\rm max}$.  

\begin{figure}[htb]
\vspace{0.4in}
\begin{center}
\includegraphics[angle=0,width=0.40\columnwidth]{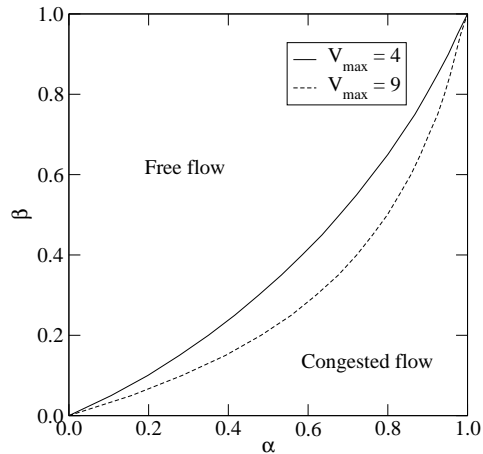}
\end{center}
\caption {Phase diagram of aggressive driving model with open boundary 
conditions for $V_{\rm max}=4$ and  $V_{\rm max}=9$.}
\label{fig-11}
\end{figure}

The phase diagram of the ADM with open boundary conditions
for $V_{\rm max}=4$ and $V_{\rm max}=9$ is shown in Fig.~\ref{fig-11}.
The system will be found either in free-flow or congested-flow regime
depending on the values of $\alpha$ and $\beta$. Here, the $\alpha$ =
$\beta$ line does not separate the free flowing and congested flow
regime. Instead, the free flow regime is larger than the congested
flow regime. The span of the free-flow regime increases with
increasing $V_{\rm max}$. In the special case $V_{\rm max}=1$ the above
insertion and extraction scheme leads us to the phase diagram of
TASEP with open boundary and where the line $\alpha$ = $\beta$
separates free flow and congested flow regime.

%%%%%%%%%%%%%%%%%%%%%%%%%%%%%%%%%%%%%%%%%%%%%%%%%%%%%%%%%%%%%%
%%%%%%%%%%%%%%%%%%%%%%%%%%%%%%%%%%%%%%%%%%%%%%%%%%%%%%%%%%%%%%

%%%%%%%%%%%%%%%%%%%%%%%%%%%%%%%%%%%%%%%%%%%%%%%%%%%%%%%%%%%
\section{Cytoplasmic dynein: from experiment to model}  
\label{sec5}
%%%%%%%%%%%%%%%%%%%%%%%%%%%%%%%%%%%%%%%%%%%%%%%%%%%%%%%%%%%

In this section we first mention the main experimentally observed 
features of the steppings of dynein motors. Then, by extending 
the ADM, we develope a simple theoretical model that captures the 
essential features of dynein stepping. 

%%%%%%%%%%%%%%%%%%%%%%%%%%%%%%%%%%%%%%%%%%%%%%%%%%%%%%%%%%%
\subsection{Stepping of dynein: experimental results}  
%%%%%%%%%%%%%%%%%%%%%%%%%%%%%%%%%%%%%%%%%%%%%%%%%%%%%%%%%%%
\label{dyneinexperiment}

In order to understand the mechanism of a single dynein motor, Mallik 
et al.\ \cite{roop} extracted the step size of single dynein motor 
from their experimental data. In their experiment, hindrance 
against forward movemnent of dyneins was caused by an opposing force. 
In principle, this hindrance could also be created by other motors. 
The smallest possible step size would be $8$ nm as the equispaced 
binding sites on the microtubule form a lattice with lattice constant 
of $8$~nm. Mallik et al.~\cite{roop} observed that in the absence of 
hindrance the step sizes of dyneins were mostly $\sim 32$~nm, i.e., 
four times the lattice constant. Moreover, the step size decreased 
with increasing hindrance: under weak hindrance the step size was 
approximately $\sim 24$~nm, under intermediate hindrance step size  
was about $\sim16$~nm, whereas under strong hindrance dynein takes 
steps of $\sim8$~nm. 
On the basis of these observations, Mallik et al.~\cite{roop} suggested 
a {\it molecular gear mechanism} for dynein motors. 

In their generic model of molecular motor traffic, Parmeggiani et al.\ 
\cite{frey} implicitly assumed a hindrance-independent step size of 
the motors. Therefore, in the light of the experimental observations 
on dynein steppings \cite{roop}, one may interpret the model 
developed in \cite{frey} to be a minimal model 
for the traffic of kinesin motors which are known to take steps of 
$8$~nm irrespective of the hindrance. Therefore, to model the traffic 
of dynein motors, which can take steps of upto $32$~nm (i.e., four 
times the lattice spacing in between the successive binding sites) 
at one go in the absence of hindrance, we need a more sophisticated 
model. In the next subsection we describe the model which we propose 
for the traffic of dynein motors.

%%%%%%%%%%%%%%%%%%%%%%%%%%%%%%%%%%%%%%%%%%%%%%%%%%%%%%%%%%%%%%%
\subsection{Dynein traffic model (DTM)}
%%%%%%%%%%%%%%%%%%%%%%%%%%%%%%%%%%%%%%%%%%%%%%%%%%%%%%%%%%%%%%%

Our dynein traffic model (DTM) has been obtained by extending the ADM 
which we have discussed extensively in the preceeding sections. 
The lattice sites in this case represent the dynein binding site on 
the microtubule track and the lattice constant is $8$~nm. In order to 
capture the fact that in the absence of hindrance a single dynein motor 
can take a single step of $32$~nm, we set $V_{\rm {max}} = 4$ in the ADM. 
Moreover, to capture the attachment and detachment of the motors from 
the microtubule track, we also allow the detachment of a motor from an 
occupied site with rate $\omega_D$ and attachment of a motor to an 
empty site with rate $\omega_A$. The state of the system is updated in 
a random sequential manner. In this DTM, a single dynein motor can 
move forward by four lattice sites (i.e., $32$~nm) in one single step 
if the available gap is greater than or equal to $32$~nm; otherwise, 
the step size will be equal to the available gap as the mutual exclusion 
between the motors hinders the motion of the following dynein.

%%%%%%%%%%%%%%%%%%%%%%%%%%%%%%%%%%%%%%%%%%%%%%%%%%%%%%%%%%%%%%%%%%%%%%
\section{Results for DTM with periodic boundary conditions}
\label{sec6}
%%%%%%%%%%%%%%%%%%%%%%%%%%%%%%%%%%%%%%%%%%%%%%%%%%%%%%%%%%%%%%%%%%%%%%

%%%%%%%%%%%%%%%%%%%%%%%%%%%%%%%%%%%%%%%%%%%%%%%%%%%%%%%%%%%%%
%\begin{figure}[tbp]
\begin{figure}[htb]
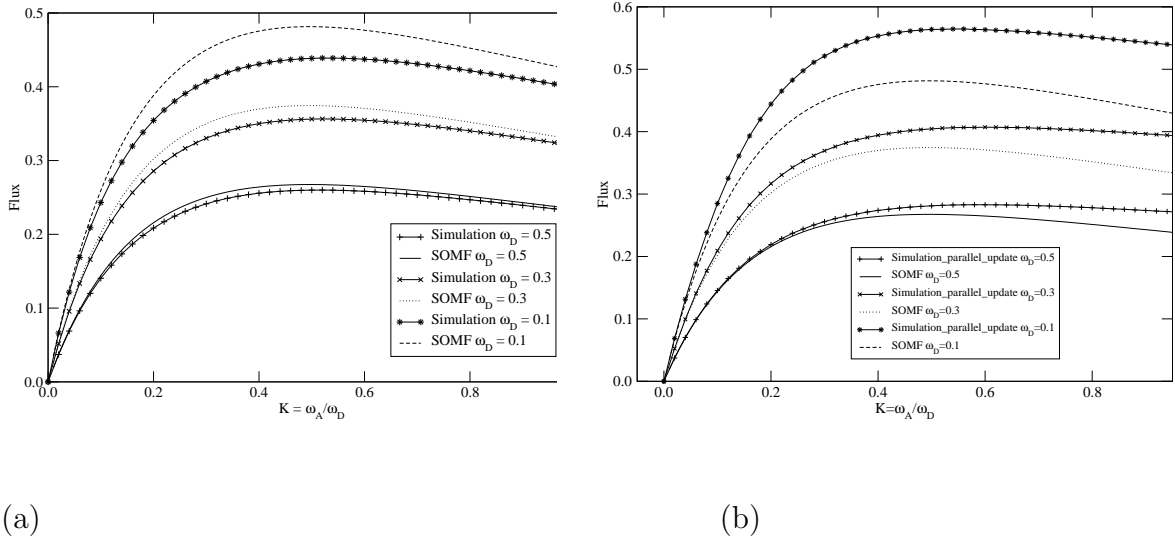

%\vspace{0.5cm}
\begin{center}
\includegraphics[width=0.49\columnwidth]{kunwarfig11a.eps}
\includegraphics[width=0.49\columnwidth]{kunwarfig11b.eps}
\end{center}
\phantom{asasasda}(a) \hspace{3.5in}(b)
\caption{Flux $f$ in the DTM with $V_{max} = 4$ for different 
values of the binding constant $K$ and with (a) random-sequential 
updating, (b) parallel updating.} 
\label{dyperiodic}
\end{figure}
%%%%%%%%%%%%%%%%%%%%%%%%%%%%%%%%%%%%%%%%%%%%%%%%%%%%%%%%%%%%%

In the case of periodic boundary condition we allow attachment and 
detachment of motors from all lattice sites with probability 
$\omega_A$ and $\omega_D$ respectively. This prescription is very 
similar to that followed by Parmeggiani et al.\cite{frey} in their 
generic model of molcular motor traffic. In order to compare and 
contrast our DTM with the model of Parmeggiani et al.\cite{frey}, 
all our computer simulations of the DTM have been carried out by 
random sequential updating which was adopted in ref.\cite{frey}. 
We have carried out limited investigation of the DTM also by 
implementing parallel updating. But, unless explicitly stated 
otherwise, by computer simulation of the DTM we shall mean simulation 
using random sequential updating. 

Fig.~\ref{dyperiodic} shows the values of flux $f$ as a function of 
the binding constant $K = \omega_A/\omega_D$ for different values 
of $\omega_D$ obtained from the computer simulations implementing 
random-sequential updating (Fig.~\ref{dyperiodic}(a)) and parallel 
updating (Fig.~\ref{dyperiodic}(b)). In both the cases the flux 
initially increase with increasing $K$. However, beyond a threshold 
value of $K$ the flux starts decreasing with the further increase 
of $K$. Comparing the figures \ref{dyperiodic}(a) and \ref{dyperiodic}(b) 
we find that, for the same set of values of the parameters, higher 
flux is obtained with parallel updating; this effect becomes more 
pronounced at higher $\omega_D$. This is consistent with the well 
known result that flux obtained with parallel updating is higher 
than that obtained with random-sequential updating even in the 
simpler situation of TASEP.

We have carried out a Site-Oriented Mean Field (SOMF) calculation 
for the DTM following the same procedure as we followed earlier for 
the ADM. For $V_{\rm {max}}=4$, we get 
\begin{eqnarray}
\label{eq-dynpar1}
{c_0}(i,t+1) &=& c(i+1,t)[{c_1}(i,t)+{c_2}(i,t)+{c_3}(i,t)+{c_4}(i,t)] \\
\label{eq-dynpar2}
{c_1}(i,t+1) &=& qc(i+1,t)d(i,t)c(i-1,t)\\
\label{eq-dynpar3}
{c_2}(i,t+1) &=& qc(i+1,t)d(i-1,t)d(i,t)c(i-2,t)\\
\label{eq-dynpar4}
{c_3}(i,t+1) &=& qc(i+1,t)d(i-2,t)d(i-1,t)d(i,t)c(i-3,t)\\
\label{eq-dynpar5}
{c_4}(i,t+1) &=& qd(i-3,t)d(i-2,t)d(i-1,t)d(i,t)c(i-4,t)
\end{eqnarray}
In the steady state, i.e. $t \rightarrow \infty$, ${c_V}(i,t)$  are
independent of $t$. For periodic boundary conditions the system 
becomes homogeneous in the steady state and hence the $i$-dependence 
of ${c_V}(i)$ also drops out. Therefore the steady state flux for 
$V_{\rm {max}}=4$ is given by
\begin{equation}
f = {c_1}+2{c_2}+3{c_3}+4{c_4} = qcd[c+2cd+3cd^2+4d^3]
\end{equation}
where $q$ is the probability that the motor does, indeed, hop, 
instead of getting detached from the track. Finally, substituting 
$q=1-\omega_D$, the steady flux in the DTM, under the SOMF 
approximation, is given by 
\begin{equation}
\label{eqn-flux}
f= cd\bigl((1-\omega_D)c+2(1-\omega_D)cd+3(1-\omega_D)cd^2+4
(1-\omega_D)d^3\bigr)
\end{equation}
where $c$ is given by the well-know ratio $K/(1+K)$ of Langmuir
equilibrium density and $d=1-c$. In Figs.~\ref{dyperiodic}(a) and 
(b) we have shown the curves obtained from Eq.~(\ref{eqn-flux}) 
to compare the predictions of the SOMF theory with the corresponding 
simulation data. Although the agreement is not very good, we 
find that the SOMF provides better estimates of flux at higher 
values of the detachment probability $\omega_D$. 

One interesting feature of the SOMF result  
is that it is an overestimate of the flux for random-sequential 
updating whereas it is an underestimate of that corresponding to 
parallel updating. Moreover, the SOMF estimates are closer to 
the simulation data from random-sequential updating than those 
from parallel updating. This behaviour indicates that the 
underlying correlations are rather subtle.

%%%%%%%%%%%%%%%%%%%%%%%%%%%%%%%%%%%%%%%%%%%%%%%%%%%%%%%%%%%%%%
\begin{figure}[htb]
\vspace{0.4in}
\begin{center}
\includegraphics[angle=0,width=0.40\columnwidth]{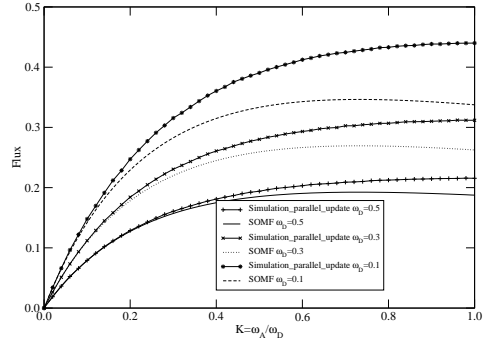}
\end{center}
\caption {The fundamental diagram in a {\it hypothetical} DTM with 
$V_{\rm max}=2$ and parallel updating under periodic boundary conditions.}
\label{hypdtm}
\end{figure}
%%%%%%%%%%%%%%%%%%%%%%%%%%%%%%%%%%%%%%%%%%%%%%%%%%%%%%%%%%%%%%
                                                                                
The agreement between SOMF theory and computer simulations is 
much better in the DTM than in the ADM. Is it an artefact of 
the different values of $V_{max}$ used in Fig.\ref{fig-7}(a)  
and Fig.~\ref{dyperiodic}(b)? In order to investigate this 
possibility, we have studied a hypothetical DTM with $V_{max} = 2$ 
by SOMF as well as by computer simulation implementing parallel 
updating. The SOMF estimate for the flux in this hypothetical 
DTM with $V_{max}=2$ is given by
\begin{equation}
f= c d (2-c) (1-\omega_d)
\end{equation}
where $c=K/(1+K)$. This SOMF estimate is compared in Fig.\ref{hypdtm} 
with the numerical data obtained from the simulation of the same 
model with parallel updating. The agreement is as good as that in 
Fig.~\ref{dyperiodic} for the actual DTM with $V_{max} = 4$.  
Thus, the reason for better success of SOMF in DTM than in ADM 
remains a challenging open problem for future investigation.

%%%%%%%%%%%%%%%%%%%%%%%%%%%%%%%%%%%%%%%%%%%%%%%%%%%%%%%%%%%%%%%%%%%%%%
\section{DTM with open boundary condition}
\label{sec7}
%%%%%%%%%%%%%%%%%%%%%%%%%%%%%%%%%%%%%%%%%%%%%%%%%%%%%%%%%%%%%%%%%%%%%%

%%%%%%%%%%%%%%%%%%%%%%%%%%%%%%%%%%%%%%%%%%%%%%%%%%%%%%%%%%%%%%
\begin{figure}[tbp]
\begin{center}
\includegraphics[width=0.6\columnwidth]{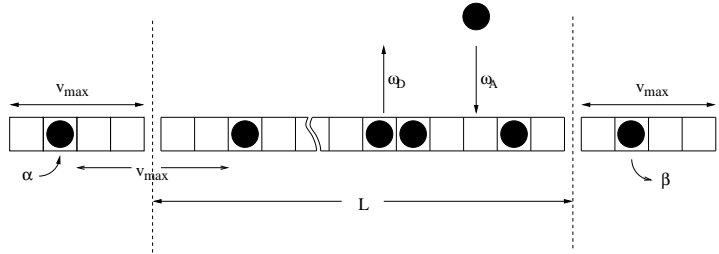}
\end{center}
\caption{Schematic representation of the DTM with open boundary conditions.} 
\label{dyneinmodel}
\end{figure}   
%%%%%%%%%%%%%%%%%%%%%%%%%%%%%%%%%%%%%%%%%%%%%%%%%%%%%%%%%%%%%%

A schematic representation of the analysed system with open boundary
conditions is given in Fig.~\ref{dyneinmodel}.  To capture the
attachment and detachment of the motors from the microtubule we allow
the detachment of a motor from an occupied site with rate $\omega_D$
and attachment of a motor to an empty site with rate $\omega_A$ in the
bulk i.e.\ from all site other except those who belong to the
reservoirs at the left and right boundary. Our proposed model with
open boundary conditions reduced to the model of Parmeggiani et al.\
\cite{frey}, which is a minimal model for the intra-cellular traffic
of kinesin motors, if one sets $V_{\rm {max}}=1$.

%%%%%%%%%%%%%%%%%%%%%%%%%%%%%%%%%%%%%%%%%%%%%%%%%%%%%%%%%%%
\begin{figure}[tbp]
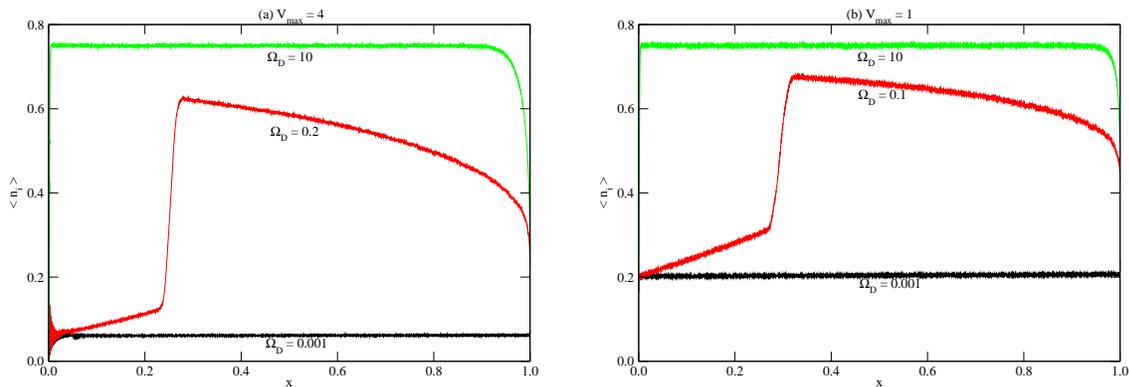

\vspace{0.1in}
\begin{center}
\includegraphics[width=0.45\columnwidth]{kunwarfig14a.eps}
\hspace{0.2in}
\includegraphics[width=0.45\columnwidth]{kunwarfig14b.eps}
\end{center}
\caption{Average density profiles $\langle n_i \rangle$ in the DTM 
obtained from Monte Carlo simulations and plotted against the 
rescaled space variable $x = i/L$ for different values of $\Omega_D$. 
The common parameter values are $L=10000, \alpha=0.2, \beta=0.6,
K=3$. } 
\label{densityprofile}
\end{figure}  
%%%%%%%%%%%%%%%%%%%%%%%%%%%%%%%%%%%%%%%%%%%%%%%%%%%%%%%%%%%

\noindent
A competition between bulk dynamics (Langmuir kinetics) and boundary
induced non-equilibrium effects (TASEP-like dynamics) is expected only
if the particles injected either at boundary or somewhere in bulk
visits a finite fraction of the total system size. In that case
particles will spend enough time on lattice to feel mutual influence
and eventually would produce collective effects.

Study of competition between bulk and boundary dynamics for 
large systems ($L \gg 1$) requires 
that the kinetic rates $\omega_A$ and $\omega_D$ decrease simultaneously
with increasing system size $L$. This can be illustrated by considering the
following heuristic argument given in ref.~\cite{frey}.
The average time $\tau$ spent by a particle before detachment is
roughly of the order of $\sim$ $1/\omega_D$. During this time $\tau$
the number of sites $n$ visited by a given particle is of the order of
$n \sim \tau$. Therefore the fraction $n/L$ ($=1/\omega_DL$) of the
lattice site visited by a given particle during this time would tend
to zero for fixed $\omega_D$ as $L \rightarrow \infty$.  In order that
a given particles explores a finite fraction of the total sites before
detaching for system size $L \gg 1$, one has to define the ``total"
detachment rate $\Omega_D = \omega_DL$ such that $\Omega_D$ remain
constant as $L \rightarrow \infty$. A similar argument shows that a
vacancy visits a finite fraction lattice sites until it is filled by
the attachment of a particle if $\omega_A$ scales to zero as
$\Omega_A/L$ with fixed ``total" detachment rate $\Omega_A$
\cite{frey}.  Therefore we define total detachment rate $\Omega_D =
\omega_DL$ and total attachment rate $\Omega_A = \omega_AL$ such that
$\Omega_D$ and $\Omega_A$ remain constant as $L \rightarrow \infty$.
Note that the binding constant $K=\omega_A/\omega_D$ remains unchanged
as $L \rightarrow \infty$.

%%%%%%%%%%%%%%%%%%%%%%%%%%%%%%%%%%%%%%%%%%%%%%%%%%%%%%%%%%%
\subsection{Density profiles in the DTM}
%%%%%%%%%%%%%%%%%%%%%%%%%%%%%%%%%%%%%%%%%%%%%%%%%%%%%%%%%%%

%%%%%%%%%%%%%%%%%%%%%%%%%%%%%%%%%%%%%%%%%%%%%%%%%%%%%%%%%%%%%%
\subsubsection{Density profiles}
\begin{figure}[tbp]
\begin{center}
\includegraphics[width=1.0\columnwidth]{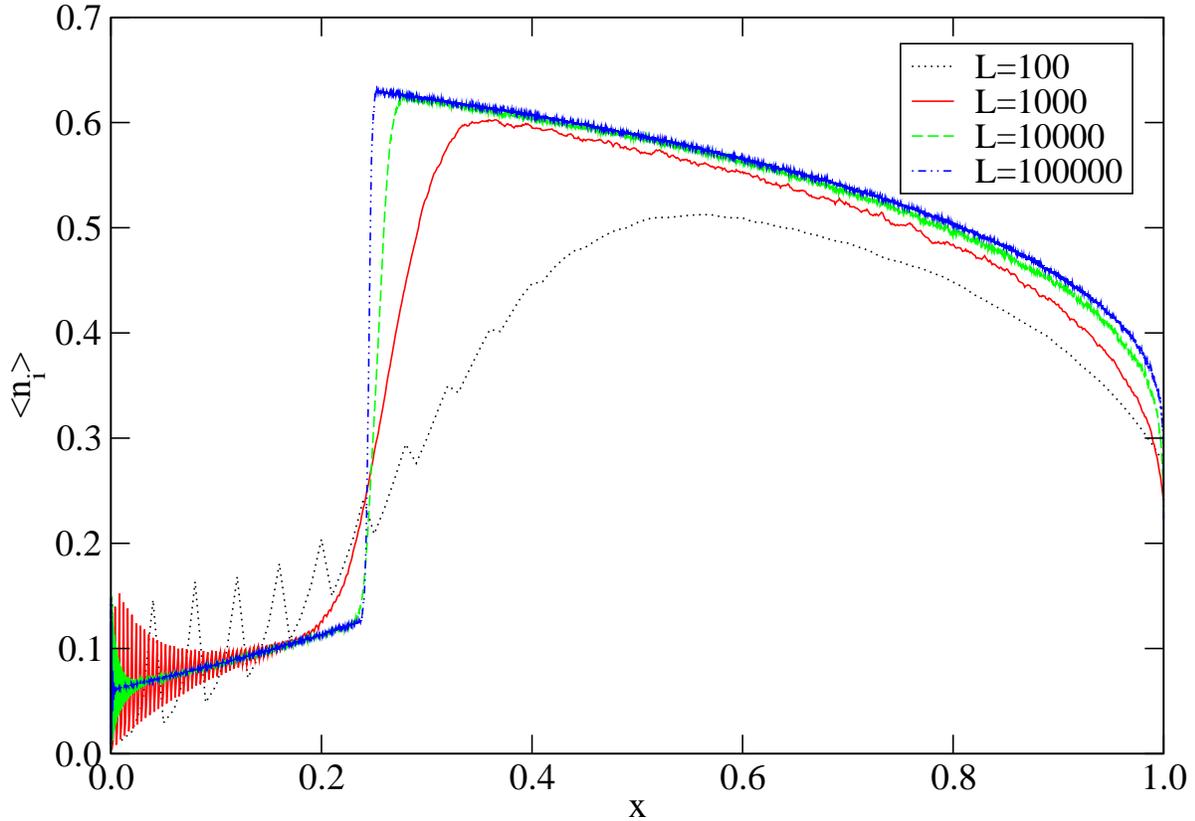}
\end{center}
\caption{Average density profiles in the DTM obtained from Monte Carlo 
simulations and plotted against rescaled space variable for different 
system sizes. The common parameter values are $\alpha=0.2, \beta=0.6,
K=3$ and $\Omega_D=0.2$.} 
\label{systemsize}
\end{figure}  
%%%%%%%%%%%%%%%%%%%%%%%%%%%%%%%%%%%%%%%%%%%%%%%%%%%%%%%%%%%%%%

In Fig.~\ref{densityprofile} (a) we have plotted the typical average
density profiles $\langle n_i \rangle$ obtained from Monte Carlo
simulations for this model in rescaled space variable $x = i/L$ (where
$n_i$ denotes occupation of nth lattice site in the bulk i.e.
$i=1,2,...L$) by choosing a path in the parameter space along the
curves with fixed $\alpha, \beta$ and $K$ while increasing $\Omega_D =
\Omega_A/K$. For low values of kinetic rates $\Omega_D, \Omega_A \ll
\alpha, \beta$ the system remain in low-density phase where average
density in the bulk remains constant. For High values of kinetic rates
$\Omega_D$ and $\Omega_A$ the system goes to high density phase where
average density in the bulk again constant.  The bulk density in this
case is determined by the well know ratio $K/(1+K)$ of Langmuir
equilibrium density. For intermediate values of the rates, for example
$\Omega_D =0.2$, the density profile in the bulk exhibits unusual
feature where regions of high density and low density are connected by
a steep rise. In Fig.~\ref{densityprofile} (b) we have shown the
density profiles obtained by Parmeggiani et al.\ \cite{frey} for their
model where $V_{\rm {max}} = 1$. Comparison of
Figs.~\ref{densityprofile} (a) and \ref{densityprofile} (b) shows that
the average bulk density in the low density phase in
Fig.~\ref{densityprofile} (a) is smaller than in
Fig.~\ref{densityprofile} (b).  This observed decrease in the bulk
density in the low density phase for identical values of the
parameters $\alpha ,\beta, K $ and $\Omega_D$ is due to the fact that
particles in low density phase move with higher velocity which leads
to higher current and lower density.  Fig.~\ref{densityprofile} (a)
also shows oscillations in the density profile at the beginning of the
system in the low density phase. These oscillations result from the
hindrance that particles have at the beginning of the system from each
other \cite{cheybanipre}.  These oscillations die out for higher
systems sites.  Comparison of Fig.~\ref{densityprofile} (a) and
Fig.~\ref{densityprofile} (b) also shows that in the case of
dynein motors the domain wall is found inside the the system for
slightly higher value of $\Omega_D$.

\begin{figure}[tbp]
\begin{center}
\vspace{0.5cm}
\includegraphics[width=0.45\columnwidth]{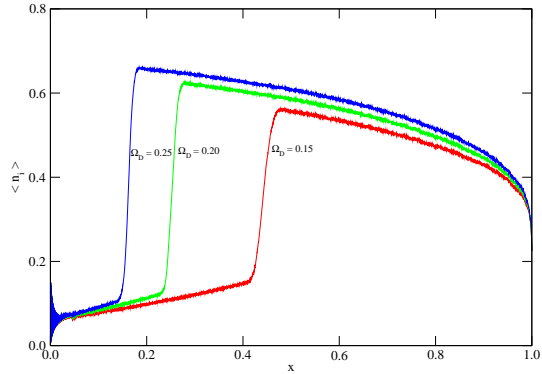}
\vspace{0.4cm}
\end{center}
\caption{Domain wall positions in the DTM for different values of 
  $\Omega_D$. The common parameter values are $L=10000, \alpha=0.2,
  \beta=0.6$ and $K=3$.}
\label{dwposition}
\end{figure}

Figure \ref{systemsize} shows the average density profile $\langle n_i
\rangle$ computed from Monte Carlo simulation in rescaled space
variable $x$ for different system sizes. The width of the transition
region decreases with increasing system size. The data obtained from
our simulations suggest a sharp discontinuity of the density profile
in terms of the rescaled space variable $x = i/L$ in the limit
$L\rightarrow \infty$. Therefore the low and high density phases
separated by a sharp domain wall coexist in our model over an
intermediate range of parameter values where boundary and bulk kinetic
rates compete against each other. This discontinuity in the density
profile is stable and the position of the domain wall is determined by
the values of the kinetic rates as shown in Fig. \ref{dwposition}.
This coexistence of high and low density phase separated by a domain
wall can be regarded as a {\it traffic jam} for molecular motors.
%%%%%%%%%%%%%%%%%%%%%%%%%%%%%%%%%%%%%%%%%%%%%%%%%%%%%%%%%%%%%
%%%%%%%%%%%%%%%%%%%%%%%%%%%%%%%%%%%%%%%%%%%%%%%%%%%%%%%%%%%%%

%%%%%%%%%%%%%%%%%%%%%%%%%%%%%%%%%%%%%%%%%%%%%%%%%%%%%%%%%%
\subsection{Phase diagrams in the DTM} 
%%%%%%%%%%%%%%%%%%%%%%%%%%%%%%%%%%%%%%%%%%%%%%%%%%%%%%%%%%

%%%%%%%%%%%%%%%%%%%%%%%%%%%%%%%%%%%%%%%%%%%%%%%%%%%%%%%%%%%%%
\begin{figure}[tbp]
\begin{center}
\includegraphics[width=1.0\columnwidth]{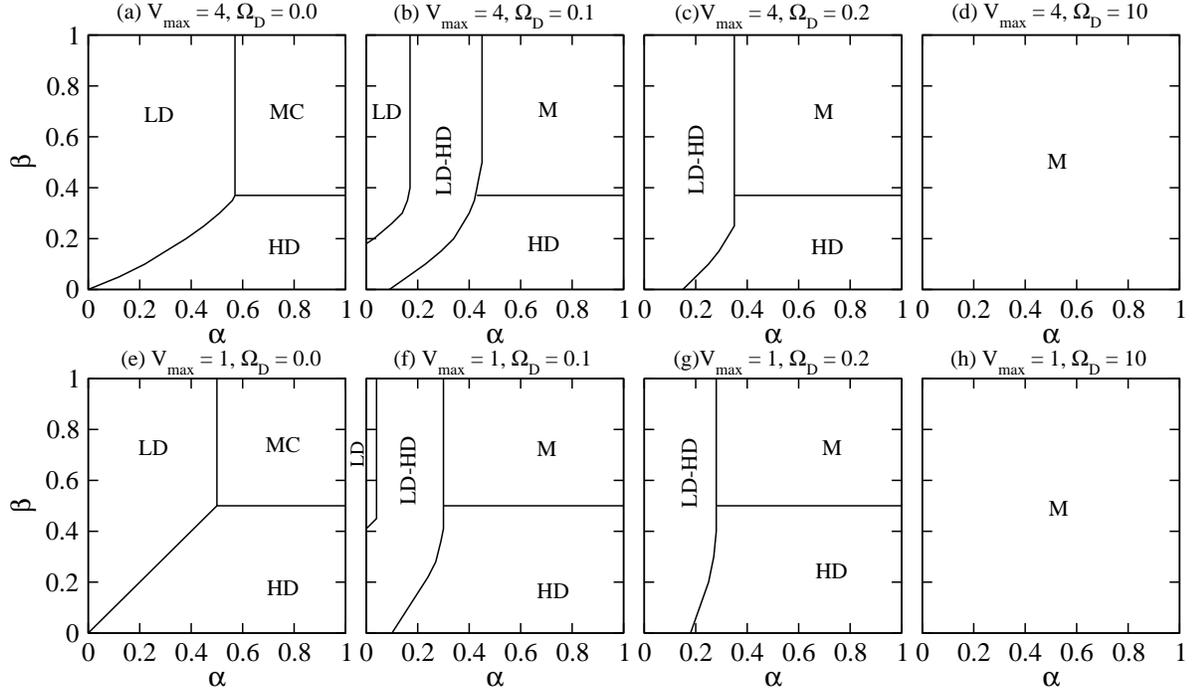}
\end{center}
\caption{Phase diagrams of the DTM for (a) $V_{\rm {max}} =4, \Omega_D =0.0$ 
(b) $V_{\rm {max}} =4, \Omega_D =0.1$ 
  (c) $V_{\rm {max}} =4, \Omega_D =0.2$ and (d) $V_{\rm {max}} =4,
  \Omega_D = 10$ (e) $V_{\rm {max}} =1, \Omega_D =0.0$ (f) $V_{\rm
    {max}} =1, \Omega_D =0.1$ (g) $V_{\rm {max}} =1, \Omega_D =0.2$
  and (h) $V_{\rm {max}} =1, \Omega_D = 10$.  Other common parameter
  values are $L=10000, K=3$.}
\label{phasediag}
\end{figure}
%%%%%%%%%%%%%%%%%%%%%%%%%%%%%%%%%%%%%%%%%%%%%%%%%%%%%%%%%%%%%

In order to identify the regions of coexistence, we have obtained the
phase diagram of our model for intra-cellular traffic of dynein motors
by varying the boundary rates $\alpha$ and $\beta$ for fixed values of
$K$ and $\Omega_D$. Fig.~\ref{phasediag}(a) shows the phase diagram for
$\Omega_D=0$.  In this case model reduces to the aggressive driving
model with random sequential updating. For $\Omega_D=0$, by varying
boundary rates $\alpha$ and $\beta$ one gets three kind of phases
namely Low density phase (LD), High density phase (HD) and Maximal
Current phase (MC). For very small values of $\Omega_D$ ($\Omega_D
\sim 0.001$) the boundary rates $\alpha$ and $\beta$ dominate and the
structure of the phase diagram is determined only by the boundary
rates $\alpha$ and $\beta$. For $\Omega_D \sim 0.001$ one obtained a
phase digram similar to Fig.~\ref{phasediag}(a). On increasing the
value of $\Omega_D$ the boundary and bulk rates start competing with
each other and in this case one gets a phase diagram where MC phase
disappears and one can identify four distinct regions in the phase
diagram namely, Low density phase (LD), Low density high density
coexistence region (LD-HD), High density phase (HD) and ``Meissner"
(M) phase. The Meissner phase  \cite{frey} has some interesting
features that are genuinely distinct from the High density phase. The
density profile in the bulk is independent of the boundary rates
$\alpha$ and $\beta$ and is determined only by the bulk. On further
increase of $\Omega_D$ the low density phase also disappears from the
phase diagram as shown in Fig.~\ref{phasediag}(c) and in this case one
gets HD phase, M phase and coexistence region. For large values of
$\Omega_D$ the phase diagram is spanned only by the M phase as shown
in Fig.~\ref{phasediag} (d).  In Fig.~\ref{phasediag}(e)-(h) we have
shown phase diagrams of the model of Parmeggiani et al.\ \cite{frey}
for the identical set of parameters. Comparison of the phase diagrams
shown in Figs.~\ref{phasediag}(a)-(d) and \ref{phasediag}(e)-(h)
shows that the regions of coexistence of Low and High density phases
are slightly different in the $\alpha - \beta$ plane for identical
values of the parameters $K$ and $\Omega_D$.

%%%%%%%%%%%%%%%%%%%%%%%%%%%%%%%%%%%%%%%%%%%%%%%%%%%%%%%%%%%%%
\subsection{Transportation efficiency}
%%%%%%%%%%%%%%%%%%%%%%%%%%%%%%%%%%%%%%%%%%%%%%%%%%%%%%%%%%%%%

Comparison of $32$ nm step of dynein at in the absence of hindrance and 
$8$ nm step of kinesin implies that as a cargo transporter the dynein 
is four times more fuel-efficient than kinesin as both require one 
molecule of ATP as fuel. To study the transportation efficiency of 
dynein motors and kinesin motors as a function of the parameters $K$ and 
$\Omega_D$ we define the transportation efficiency by the relation 
\begin{equation}
  \% \textnormal{Transportation efficiency} = \frac{\textnormal{No. of
      steps taken}} {\textnormal{No. of attempts made}\times4} \times
  100
\end{equation}

%%%%%%%%%%%%%%%%%%%%%%%%%%%%%%%%%%%%%%%%%%%%%%%%%%%%%%%%%%%%%%%%%
\begin{figure}[tbp]
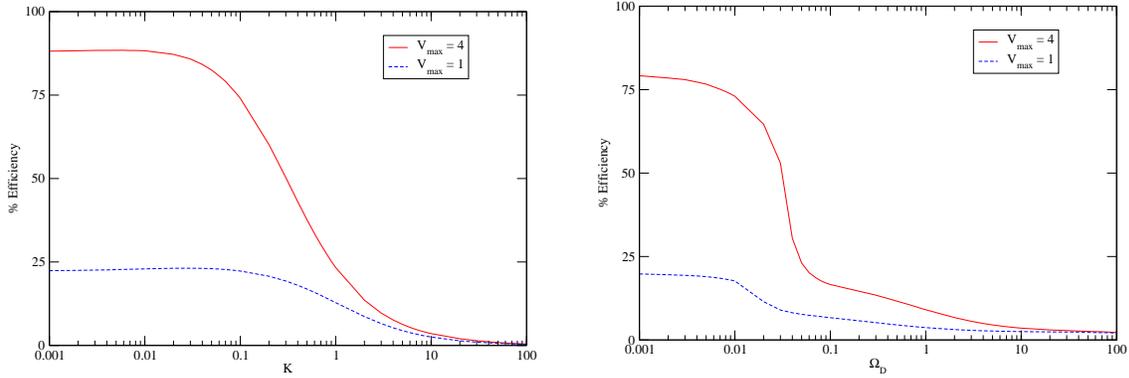

\begin{center}
\includegraphics[width=0.45\columnwidth]{kunwarfig18a.eps}
\hspace{0.2in}
\includegraphics[width=0.45\columnwidth]{kunwarfig18b.eps}
\end{center}
\caption{Efficiency of dynein and kinesin motors for (a) different 
  values of $K$ and (b) $\Omega_D$. The common parameter values are
  $L=1000, \alpha=0.2$ and $\beta=0.6$}
\label{efficiency}
\end{figure}  
%%%%%%%%%%%%%%%%%%%%%%%%%%%%%%%%%%%%%%%%%%%%%%%%%%%%%%%%%%%%%%%%%

The above relation has been defined in such a way that if a dynein
motor takes 4 steps of 8~nm in each attempt (i.e. for each ATP
hydrolysis) then its efficiency will be 100 \% similarly if a kinesin
motor takes 1 step of 8~nm in each attempt then its efficiency will be
equal to 25 \%. The transportation efficiency of cytoplasmic dynein
motors ($V_{\rm {max}} = 4$) and kinesin motors ($V_{\rm {max}} = 1$)
are plotted in Fig.~\ref{efficiency} (a) and Fig.~\ref{efficiency} (b)
for different values of $\Omega_D$ and $K$. There is practically no
difference between the efficiency of dynein and kinesin motors for
very large values of $\Omega_D$ and $K$ ($\Omega_D \sim 10$ and $K
\sim 10$) as for very large values of $\Omega_D$ and $K$ system is
found in high density phase (HD).

%%%%%%%%%%%%%%%%%%%%%%%%%%%%%%%%%%%%%%%%%%%%%%%%%%%%%%%%%%%%%%%%%%%%%%%%%%%%%%
\section{Summary and conclusions}
\label{sec8}
%%%%%%%%%%%%%%%%%%%%%%%%%%%%%%%%%%%%%%%%%%%%%%%%%%%%%%%%%%%%%%%%%%%%%%%%%%%%%%

In this paper we have first investigated the properties of the 
aggressive driving model (ADM) which is a simple cellular automata 
model for vehicular traffic. One of the motivations for considering 
this model is that the rule for aggressive driving can be naturally 
extended to capture the special features of step sizes of dynein 
motors and, therefore, the ADM is ideally suited for extending so as 
to study intra-cellular molecular motor traffic by dynein motors. 

The ADM shows different behavior for $V_{\rm max} = 1$ and 
$V_{\rm max} > 1$. For $V_{\rm max} = 1$ the model is identical
to the NaSch model with $V_{\rm max} = 1$ which has perfect 
particle-hole symmetry. This symmetry is broken for $V_{\rm max} > 1$. The 
fundamental diagram of this model in the special limit 
$V_{\rm max} = \infty$ has a form which is quite different from that 
of the NaSch model in the limit $V_{\rm max} = \infty$. We have also 
shown few distance headway and time-headway distributions. We have 
calculated the fundamental diagram using two different mean-field 
approaches, namely, site-oriented mean-field approach (SOMF) and 
car-oriented mean-field approach (COMF). A simple SOMF theory shows 
a poor agreement with the simulation data. However, an improved mean 
field theory, namely COMF, shows good agreement with the numerical 
data obtained from computer simulations. We compare our ADM with the
Nagel-Schreckenberg model which captures essential features of normal
driving.  We have also investigated the density profiles and phase 
diagrams of this model replacing the periodic boundary conditions 
by open boundary conditions. The density profile of this model with 
open boundary conditions shows periodic structures in the free-flowing
regime whose period of oscillation depends on the maximum attainable
velocity $V_{\rm max}$. 

We have extended the ADM to develope a dynein traffic model (DTM) 
which is a model of intra-cellular molecular motor traffic from 
cell periphery towards the nucleus of the cell. We have investigated
the properties of this model with periodic and open boundary conditions. 
Under open boundary conditions, DTM shows an unusual feature where 
low and high density phases separated by a static domain wall coexist 
over a range of parameter values which can be interpreted as a traffic 
jam of molecular motors. This is in sharp contrast to the phase 
diagram of the ADM which does not exhibit such coexistence of 
congested and free-flowing regions. The occurrence of the phase is, 
thus intimately related to the competition between the hopping and  
the kinetics of attachment/detachment of the motors on the track. 
Finally, we have compared the efficiencies of dynein and kinesin motors 
for different values of parameters. For very large values of the 
parameters $\Omega_D$ and $K$ system is found in the high density 
phase and, in that case, one observes practically no difference between 
the efficiencies of kinesin and dynein motors.

To our knoweledge, our DTM is the first model of traffic-like collective 
transport of {\it dynein} motors on filamentary microtubule tracks. A 
model that incorporates both the species of dyneins and kinesin motors, 
which move in opposite directions along the same track, may provide 
deep insight into experimentally observed bidirectional traffic 
\cite{gross}. 

\vspace{0.3cm}

%%%%%%%%%%%%%%%%%%%%%%%%%%%%%%%%%%%%%%%%%%%%%%%%%%%%%%%%%%%%%%%%%%%
\noindent{\bf Acknowledgements}: This work has been supported (through 
DC), in part, by the Council of Scientific and Industrial Research 
(CSIR) of the government of India. DC also thanks Max-Planck Institute 
for Physics of Complex Systems, Dresden, and Prof. Frank J\"ulicher 
for hospitality during a short visit when a part of this manuscript 
was prepared.
%%%%%%%%%%%%%%%%%%%%%%%%%%%%%%%%%%%%%%%%%%%%%%%%%%%%%%%%%%%%%%%%%%%

%\bibliographystyle{plain}
%\begin{thebibliography}{1}
\section*{References}

%%%%%%%%%%%%%%%%%%%%%%%%%%%%%%%%%%%%%%%%%%%%%%%%%%%%%%%%%%%%%

\end{document}